\documentclass[preprint]{aastex}

\newcommand{\ie}{{\frenchspacing\it i.e.}}

\newcommand{\etal}{{\frenchspacing\it et al. }}

\newcommand{\lsim}{\hbox{ \rlap{\raise 0.425ex\hbox{$<$}}\lower
0.65ex\hbox{$\sim$} }}
\newcommand{\gsim}{\hbox{ \rlap{\raise 0.425ex\hbox{$>$}}\lower
0.65ex\hbox{$\sim$} }}

\shorttitle{NoSOCS IV: Intermediate-$z$ Clusters}
\shortauthors{Lopes et al.}

\begin{document}

\title{The Northern Sky Optical Cluster Survey IV: An Intermediate Redshift
Galaxy Cluster Catalog and the Comparison of Two Detection Algorithms \\
}

\author{P.A.A. Lopes\altaffilmark{1}, R.R. de Carvalho\altaffilmark{2,1},
R.R. Gal\altaffilmark{3}, S.G. Djorgovski, S.C. Odewahn\altaffilmark{4},
A.A. Mahabal, R.J. Brunner\altaffilmark{5}}

\affil{Palomar Observatory, Caltech, MC 105-24, Pasadena, CA 91125}

\altaffiltext{1}{Current address: Instituto Nacional de Pesquisas Espaciais - Divis\~ao de Astrof\'isica,
Avenida dos Astronautas, 1758, S\~ao Jos\'e dos Campos, SP 12227-010, Brasil \\
{\indent Email: paal@das.inpe.br} }
\altaffiltext{2}{Current address: Observat\'orio Nacional, Rua General Jos\'e Cristino, 77,
 Rio de Janeiro, RJ 20921-400, Brasil}
\altaffiltext{3}{Current address: UC Davis, Dept. of Physics, One Shields Ave, Davis, CA 95616}
\altaffiltext{4}{Current address: Hobby-Eberly Telescope, HC75 - Box 1337-10, Ft. Davis, TX 79734-5015}
\altaffiltext{5}{Current address: University of Illinois, Dept. of Astronomy, 1002 W. Green St., Urbana, IL 61801}

\begin{abstract}

We present an optically selected galaxy cluster catalog from $\sim$
2,700 $\square^\circ$ of the Digitized Second Palomar Observatory Sky
Survey (DPOSS), spanning the redshift range $0.1 \lesssim z \lesssim
0.5$, providing an intermediate redshift supplement to the previous
DPOSS cluster survey. This new catalog contains 9,956 cluster candidates
and is the largest resource of rich clusters in this redshift range to
date.  The candidates are detected using the best DPOSS plates based
on seeing and limiting magnitude. The search is further restricted to
high galactic latitude ($|b| > 50^\circ$), where stellar
contamination is modest and nearly uniform.  We also present a
performance comparison of two different detection methods applied to
this data, the Adaptive Kernel and Voronoi Tessellation techniques.
In the regime where both catalogs are expected to be complete, we find
excellent agreement, as well as with the most recent surveys in the
literature.  Extensive simulations are performed and applied to the
two different methods, indicating a contamination rate of $\sim 5 \%$.
These simulations are also used to optimize the algorithms and
evaluate the selection function for the final cluster
catalog. Redshift and richness estimates are also provided, making
possible the selection of subsamples for future studies.

\end{abstract}


\section{Introduction}

Clusters of galaxies constitute the largest bound structures in the universe.
Hence, clusters have been widely used to trace the mass distribution
and its evolution
\citep{bs83, phg92, guz92}, as well as to place constraints on cosmological
models \citep{bah97, evr89, via96, cal97}.
Galaxy clusters are also likely composed of coeval stellar systems, and are
therefore well suited for studying the formation and evolution of galaxies
in dense environments \citep{ara93, bo84, dre97, mar01}. Studies
of this nature require the use of a well defined and understood sample of
galaxy clusters. Although whole sky cluster catalogs exist, they typically
are subjective in nature or do not span a large redshift range. On the other
hand, deep cluster catalogs are limited to small regions of the sky.

In the last fifteen years many authors have used different wavelengths
and techniques to identify clusters of galaxies. Each method has its
own biases, and an ideal sample would be drawn from a combination of
data sets and techniques. Optical cluster catalogs use galaxy
overdensities as a proxy for mass overdensities.  Cluster-scale mass
condensations lacking a significant galaxy population are not likely
to be recovered by optical surveys, although the existence of many
such clusters is questionable. Optical imaging catalogues are
observationally inexpensive, but projection effects are a critical
drawback. However, such effects can be minimized with the aid of
colors. X-ray catalogues do not suffer from projection effects, but
are biased against clusters with an unresolved gas distribution or
lower gas content.  Some other cluster detection methods include
weak-lensing \citep{wit01} and the search for distortions in the
cosmic microwave background, where clusters scatter the microwave
background radiation via the Sunyaev-Zeldovich effect \citep{sun80}.

Most cluster studies to date have made use of the Abell catalog
\citep{abe58, aco89}, which is the result of visual inspection of
photographic plates. Other examples of subjective catalogs generated
from plate data can be found in \citet{zwi68} and \citet{gun86}. The
main drawbacks of human-based cluster searches are incompleteness,
lack of reproducibility, and lengthy generation time, as well as
difficulty in quantifying the selection effects associated with
the resulting catalog.

\citet{she85} was the first to use an automated method to search for
clusters, with much progress since that time. Many automated cluster
finding techniques have been developed, with some applied to wide
field data. A few examples are the APM Cluster Survey \citep{dal97},
the EDSGC \citep{lum92} and the EDCCII \citep{bra00}. Except for the
work of \citet{gun86} all the catalogs mentioned above sample the
nearby universe. Automated or not, all of these surveys are based on
plate data, including that of \citet{gal03}, which contains rich
clusters to $z\sim0.3$ in comparison to the $z\sim0.15-0.20$ limit of
previous surveys.

A large number of deep optical/near-IR surveys have recently become
available, reaching as deep as $z\sim1.4$,
but their sky coverage is at most a few tens of square degrees
The pioneering work is
that of \citet{pos96}, who developed and utilized a matched filter
technique. Following this work, a large number of cluster catalogs
have become available both in the northern and southern hemispheres
\citep{ols99, lob00, gla00, gon01, pos02}.  At $0<z<0.6$ there are
also preliminary results from the Sloan Digital Sky Survey (SDSS),
such as those of \citet{kim01}, \citet{ann99} and \citet{got02}.

A necessary byproduct of these surveys was the development of various
techniques for cluster detection, which utilize different properties
of the clusters.  These algorithms and techniques include the
matched filter (MF) \citep{pos96} and all its variants \citep{kaw98,
kep99, lob00, sch98, kim02}, Voronoi Tessellation \citep{ram01,
kim02}; the adaptive kernel (Gal et al. 2000, 2003, 2004b - hereafter
Papers I, II and III respectively), surface brightness fluctuations in
shallow images \citep{gon01}, and methods based on the color and
magnitude of the cluster galaxy population \citep{gla00, got02,
ann99}. Nonetheless, only a few authors have compared the performance
of different cluster search techniques on the same data. \citet{kim01}
and \citet{kim02} applied a matched filter, an adaptive matched filter
(AMF) and a Voronoi Tessellation technique (VT) with a color cut to
$\sim 152$ $\square ^\circ$ of SDSS commissioning data. The final
comparison merged the MF and AMF in a hybrid matched filter (HMF)
which uses the MF to create likelihood maps and detect cluster
candidates, while the AMF determines richness and redshift more
precisely, resulting in a final comparison with two cluster detection
algorithms (HMF and VT). \citet{got02} compared their results with
\citet{kim01} and \citet{ann99}.  \citet{bah03} compared the results
from the HMF and the maxBCG technique of \citet{ann99}.

This paper presents a galaxy cluster catalog which is an extension of
those presented in papers II and III. It covers 2,700 $\square
^\circ$ of DPOSS data and is expected to be complete for rich clusters
out to $z\sim0.3$, with clusters still detected to $z\sim0.5$. In the
regime of poor systems, this catalog is shallower than the SDSS
preliminary cluster catalogs \citep{kim02, got02, ann99, bah03}, but
the sky coverage is larger.  The main goal of this project is to
provide a catalog of rich structures to $z \sim 0.5$ covering the
northern hemisphere, with a low contamination rate at the
level of $\sim 5\%$.  As no other similar resource is available to
date, this catalog represents a valuable reference for follow-up
studies.  Our catalog will also serve as a valuable cross reference
for the final SDSS cluster catalog as well as X-ray based catalogs.

We also demonstrate the feasibility of using photometric survey data
for cluster detection in the regime where star/galaxy separation is
not reliable. At the faint end of DPOSS ($m_r > 19.5$) misclassified
stars contaminate the galaxy catalogs. As in \citet{pos02} (hereafter
P02), we assess the probability that a faint object classified as a
star is indeed a galaxy. We apply this statistical correction as a
function of magnitude to obtain an object catalog in agreement with
the galaxy counts to $m_r=21.1$.  In this way, we minimize stellar
contamination effects which would otherwise reach 25\% of the galaxy
counts.  We also restrict the search to fields at high galactic
latitude, where the star counts are approximately uniform. Finally, as
a byproduct, this work presents a performance comparison of two
different cluster search algorithms applied to DPOSS data.

Section 2 gives a brief overview of DPOSS, with emphasis on the
limitations imposed by DPOSS data on this study, and the feasibility
of exploiting it for cluster detections to $z \sim 0.5$. $\S$3
outlines the details of the two methods employed here: Voronoi
Tessellation (VT) and the Adaptive Kernel (AK). $\S$4 describes the
simulations used to optimize our algorithms and evaluate the
contamination rate. In $\S$5 we present redshift and richness
estimates for the cluster candidates, as well as the catalogs derived
by each method. In $\S$6 we evaluate the selection function for each
algorithm. In $\S$7 we perform a detailed comparison of the results
obtained using the two methods, and in $\S$8 we compare our catalog to
previous surveys. Our final results are summarized in $\S$9. We adopt
a cosmology where H$_0$ = 67 km s$^{-1}$ Mpc$^{-1}$ and q$_0$ = 0.5
throughout.  For reference, the difference in distance modulus between
this and the currently favorable cosmology with $h = 0.72$, $\Omega_m
= 0.27$ and $\Omega_{\Lambda} = 0.73$ is 0.049 mag at $z = 0.2$, and 0.265
mag at $z = 0.5$, in the sense of fainter magnitudes for the
concordance cosmology. These differences would affect our results only
for the lowest contrast systems (generally poor or distant clusters).

\section{DPOSS Data}

The Second Palomar Observatory Sky Survey (POSS-II) \citep{rei91} comprises
894 fields covering the entire northern sky ($\delta > -3^\circ$).
Each field covers an area of $6.5^\circ$ x $6.5^\circ$ with
overlapping regions of $\sim1.5^\circ$. Plates were taken in 3
bands: blue-green (IIIa-J + GG395, $\lambda_{eff} \approx$ 4800{\AA}),
red (IIIa-F + RG610, $\lambda_{eff} \approx$ 6500{\AA}) and
very near-IR (IV-N + RG9, $\lambda_{eff} \approx$ 8500{\AA}).
Those are digitized at STScI, using modified PDS scanners.
The digitized data have a 1$''$ pixel scale.
After processing and calibrating all the scans at Caltech, each 
plate results in a catalog with $\sim$ 60 attributes
measured per object.

A large amount of CCD  data was taken for photometric calibration
purposes \citep{gal04a} bringing the final magnitudes to the Gunn
system. Typical limiting magnitudes are $g_J \sim 21.5,
$ $r_F \sim 21.0,$ and $i_N \sim 19.5$. The photometrically 
calibrated data has typical {\it rms} photometric errors of 
0.25$^m$ at $m_r$ = 19.5, reaching 0.4$^m$ at $m_r$ = 21. The mean 
zero-point error is negligible, although it has a 0.07$^m$ 1$\sigma$ 
scatter, with little magnitude dependence in the $r$-band. Star-galaxy 
separation is accurate to the 90\% level at $m_r < 19.5$ \citep{ode04}. 
More details of the survey can be found in \citet{djo04}.

\subsection{How far can we see with DPOSS?}

To detect clusters at a given redshift, a photometric survey should be deep
enough to sample sources significantly fainter than the characteristic
luminosity at that redshift. Figure 1 shows the apparent
magnitude-redshift relation for elliptical galaxies at different
ranges around $m^*$. We use $M^*_r = -21.52$ \citep{pao01} and
assume clusters are dominated by early-type galaxies.
The K-correction for elliptical galaxies
is derived with the use of the spectral energy distributions from
\citet{col80}, convolved with the DPOSS $r$-band filter.
We adopt $m_r=21.1$ as the limiting magnitude of the
survey, as indicated by the vertical line in Figure 1. If we assume
that we need to go as deep as $m^*+1$ to detect a significant
number of elliptical galaxies to identify a cluster over the
background, then the survey limit lies below $z = 0.4$.
At $z \sim 0.5$ we are able to identify only the richest systems,
which have a large number of galaxies brighter than $m^*$. We should keep
in mind that these limits are valid for clusters composed entirely
of early-type galaxies. The presence of late-type galaxies would render
distant cluster members brighter as the K-correction effects are less
strong.

Figure 2 shows the differential magnitude distribution of galaxies
within the projected area of four rich galaxy clusters (solid lines)
with known spectroscopic redshifts and richnesses given by N$_{gals}$
(\S5.2).  Field counts are also shown (dotted lines). All counts are
normalized to one square degree. The shift of the luminosity function
to fainter apparent magnitudes as we go to higher redshifts is
obvious.  However, it is clear from this figure that the cluster
profile is easily differentiated from the background counts for these
rich clusters to at least $z \sim 0.35$, being still over the background
for $z \sim 0.5$ rich systems. Considering these two figures, along 
with the left panels of Figures 12 and 13 (where we show that the mean 
magnitude is generally well determined at $z \sim 0.5$) we, adopt 
$z \sim 0.5$ as a formal limit for the detection of rich clusters for
the survey.

\subsection{Field Selection}

As mentioned above, the star-galaxy separation accuracy
drops below the 90\% level at $m_r > 19.5$. Photometric errors and
observing conditions also play a key role in the generation of an input
catalog for our cluster search. To mitigate these effects, we apply a
galactic latitude cut when selecting the DPOSS fields
to be used. Additionally, we exclude fields based on their limiting
magnitudes and seeing, as outlined below.

The plate catalogs used in this paper contain galaxies with $16 \le
m_r \le 21.1$.  Initially, we ignored classification for
faint objects ($m_r > 19.5$), assuming that stars would add a
constant background to the galaxy distribution at high galactic
latitudes. However, this contribution is not expected to be mild. We
therefore adopt a statistical approach to check the probability that a
faint star is indeed a galaxy. This procedure is explained below.

The first criterion to select DPOSS fields is the galactic
latitude ($|b|$). We look for regions on the sky where density gradients
as a function of $|b|$ are minimized. In these areas we
assume that the stellar contamination (regardless of the actual level)
represents a uniform background added to the galaxy distribution.
Figure 3 shows the stellar (left panel) and galaxian (right panel) number 
counts for different magnitude bins as a function of galactic 
latitude ($|b|$). The magnitude bins are indicated in the right panel.
All lines represent spline fits to the mean density from 375 DPOSS plates.
There is  a strong dependence 
on galactic latitude, mainly for low-$|b|$ fields ($|b| < 45^{\circ}$), due 
to misclassified stars close to the galactic plane, with the increase
in density due to stellar contamination in these low-$|b|$ fields.

Another issue is whether the Poissonian fluctuations in the number of
stars misclassified as galaxies could affect the cluster detection.
In order to estimate the magnitude of this effect, we computed the
expected $\sqrt{N}$ fluctuations in the number of such misclassified stars
within a typical cluster aperture, as functions of both magnitude and galactic
latitude.  We find that for $m_r < 19$, the contamination is
less than 0.1 star per cluster aperture,
increasing to $\sim 2$ stars per cluster aperture at
$m_r = 20.5$. Thus, we conclude that the effect of such fluctuations 
is negligible for our purposes.

Based on Figure 3, we conservatively select only fields with $|b| >
50^\circ$.  This is motivated by the fact that in this latitude range
we expect any contamination caused by stars to be approximately
uniform. Nevertheless, we should still avoid high contamination
levels.  Initially, we decided to simply ignore classification for
magnitudes fainter than $m_r = 19.5$ (the star-galaxy classification
limit).  Unfortunately, in this magnitude range galaxy counts do not
exceed star counts by a large margin. At $m_r = 19.5$ the ratio of
galaxy to star counts is expected to be only 1.3, increasing to 2.5 at
$m_r = 20.5$. If we ignore classification for objects with $m_r >
19.5$, 30\% of our catalog would consist of stars (misclassified as
galaxies) at these magnitudes. Since $\sim$ 84\% of all objects with
$m_r<21.1$ have $m_r > 19.5$, the overall expected stellar
contamination would be $\sim25\%$. To reduce this effect and generate
object catalogs with statistically reliable galaxy counts, we opted
for a statistical approach to assess the probability that faint stars
are actually galaxies. This methodology was previously employed by P02
and \citet{pos98}.  The procedure consists of the extrapolation of the
bright star counts ($m_r < 18.5$) to fainter magnitudes. We can then
compare the number of stars that should be in each magnitude bin to
the actual value found in the DPOSS survey, computing the probability
that a given star in a given magnitude bin is actually a misclassified
galaxy.  This function is applied to the DPOSS faint stars ($m_r >
18.5$), having as a final product an object catalog with statistically
reliable galaxy counts.  The probability function (described below) is
given by
$$P = 5.16 - 3.2 \times 10^{-3}r^2 + 4.97 \times 10^{-5}r^4 \eqno(1)$$
This effect is illustrated in Figure 4, where we plot the number counts for
stars (squares) and galaxies (triangles) as a function of magnitude, to
$m_r = 20$, in bins of $0.5^m$. The dotted line is the best
fit to the bright galaxy counts, extrapolated to $m_r = 21$. The dashed
line is the relation used to assess the probability given by equation (1);
it is the best fit to the bright star counts ($m_r < 18.5$) extrapolated
to $m_r = 21$. Equation (1) is derived from the comparison of the
number of faint stars found in DPOSS to the expected value given by
the dashed line relation. The corrected galaxy counts are shown
as circles. These are only slightly higher than the
observed counts (triangles) at $m_r = 19.5$, in agreement with
the 90\% galaxy success rate at this magnitude. The circles are also
well fit by the dotted line at faint magnitudes, demonstrating that
the corrected counts are in agreement with the extrapolated galaxy counts
from the brighter bins. 

We apply two additional criteria to select DPOSS fields for this
survey.  First, we restrict the cluster search to fields with good
seeing.  We use the intensity-weighted second moment of the light
distribution on the red plate for stellar objects (IR2$_r$) as an
indicator of the quality of the observations. Figure 5 shows the
IR2$_r$ distribution for $|b| > 50^{\circ}$ fields. We use only fields
with IR2$_r$ $\le 1.98''$ (close to the median value), as indicated by
the vertical line in the plot.  This is equivalent to seeing measures
varying from 2.0$''$ to 2.5$''$. The poor seeing is a convolution
of the atmospheric seeing, telescope optics, and the plate
scanning process.

Finally, we exclude fields with limiting magnitudes brighter than
$m_r = 21$. The top panel of Figure 6 shows the $r$ band magnitude
distribution for a typical DPOSS field. As in \citet{pic91},
we consider the magnitude limit to be the magnitude where
the distribution starts to drop steeply (as indicated by the arrow).
The bottom panel of the same figure shows the distribution
of magnitude limits, with the chosen cut indicated by
a vertical line at $m_r = 21$.

To summarize, we select the best fields for this project based on
three criteria. Initially we begin with 375 fields. Excluding fields
with poor photometry and at $|b| < 50^\circ$ leaves 146 fields, which is
reduced to 109  after selecting those with IR2 $\le 1.98''$.
Finally we eliminated one field which has a bright magnitude
limit, $m_r < 21$. The final sample used for the cluster search thus
comprises 108 fields distributed over the northern sky, providing an
area coverage of $\sim 2,700$ square degrees. 

\section{Cluster Detection Algorithms}

In Papers I, II and III we detected galaxy clusters using the DPOSS
galaxy catalogs restricted to $m_r \le 19.5$. At this limit, all real
objects are detectable in both the $r$ and $g$ bands, so we require
that all objects to $m_r = 19.5$ must also have a counterpart in the
$g$ band, to avoid spurious detections associated with satellite or
airplane trails and plate defects.  For the deeper survey presented
here, this requirement is no longer practical. At the current limit of
$m_r = 21.1$ there are $J$ (blue) plates which are not as deep as the
$F$ (red) plates, resulting in real objects detected only in the
$r$-band which have no counterpart in the shallower $g$-band
catalogs. We have therefore opted to use only the $r$-band when
preparing the galaxy catalogs.  As described in \citet{djo04} this
results in $\sim10\%$ of $r$-band only detections being spurious;
these are mainly associated with meteor and aircraft trails. We
exclude such objects by fitting a linear relation to trails present in
our galaxy catalogs and removing detections in the corresponding
areas. Approximately 20\% of plates require this special treatment. An
example galaxy catalog for a plate with a cleaned trail (in the bottom
right corner) is plotted in Figure 7. As in the previous papers,
detections in the vicinity of bright objects (where our photometry
fails) are also excised.

Another difference from the previous papers is the use of only the central
part of each plate. We apply cuts in pixel coordinates, using only objects
with 2000 $\le X,Y \le$ 21000 (out of the maximum 0
$\le X,Y \le$ 23040), resulting in catalogs of typically $\sim$ 116000
galaxies over an area of $\sim$ 25 square degrees (compared
to 34 square degrees from Papers II and III). This is necessary to avoid the
heavily vignetted areas near the plate edges. The mean galaxy density
is $4.6\times10^3$ galaxies per square degree (approximately six
times higher than in Papers II and III). The survey covers
$\sim 2,708 \square^{\circ}$, containing $\sim 1.3 \times10^7$ galaxies.

In this paper two independent techniques are employed to detect
galaxy clusters. One, the adaptive kernel technique (AK) \citep{sil86} was
already used by our group (Papers I, II and III), while the second technique
uses Voronoi Tessellation (VT). The main differences between the  applications
 of these two methods to DPOSS data are:

\begin{description}
\item[(i)]The AK is applied only to the projected distribution of
galaxies, while the VT method also incorporates magnitude information.

\item[(ii)] As our goal in the present work is the detection of
intermediate redshift rich clusters, the galaxy catalogs used by the
AK contain only objects with 19 $\le m_r \le 21.1$. We expect to
find clusters with luminosity functions spanning these fainter
magnitudes and thus avoid low-$z$ clusters.  Initially we planned to
employ the VT technique using the same galaxy catalogs, but because we
can use magnitudes to minimize background/foreground contamination, we
opted for a brighter cut at $m_r = 16$, instead of $m_r = 19$, as
detailed in section 3.1.
\end{description}

The next two sections present a brief description of both techniques
applied to DPOSS data, with further detail in the references
provided below.

\subsection{The Voronoi Tessellation Technique}

Considering a homogeneous distribution of particles it is possible to
define a characteristic volume associated with each
particle.  This is known as the Voronoi volume, whose radius is of the
order of the mean particle separation.  Voronoi Tessellation has been
applied to a variety of astronomical problems. A few examples are
found in \citet{ikt91}, \citet{zan95}, \citet{ela96},
\citet{dor97}. \citet{ebw93} used Voronoi Tessellation to identify
X-ray sources as overdensities in X-ray photon counts.  \citet{kim02}
and \citet{ram01} looked for galaxy clusters using Voronoi
Tessellation (VT). As pointed out by \citet{ram01} one of the main
advantages of employing VT to look for galaxy clusters is that this
technique does not distribute the data in bins, nor does it assume a
particular source geometry intrinsic to the detection process.  The
algorithm is thus sensitive to irregular and elongated structures.

The parameter of interest in our case is the galaxy density. When
applying VT to a galaxy catalog, each galaxy is considered as a seed
and has a Voronoi cell associated to it. The area of this cell is
interpreted as the effective area a galaxy occupies in the plane. The
inverse of this area gives the local density at that point. Galaxy
clusters are identified by high density regions, composed of small
adjacent cells, \ie, cells small enough to give a density value higher
than the chosen density threshold.  An example of Voronoi Tessellation
applied to the DPOSS field in Figure 7 is presented in
Figure 8. For clarity, we show only galaxies with $17.0 \le m_r \le
18.5$.

In order to detect galaxy clusters using Voronoi Tessellation we use
the code employed by \citet{ram01}. It uses the {\it triangle} C code
by \citet{shw96} to generate the tessellation. The code is designed to
avoid the borders of the field, as well as the excised areas around
saturated objects. The algorithm identifies cluster candidates based
on two primary criteria.  The first is the density threshold, which is
used to identify fluctuations as significant overdensities over the
background distribution, and is termed the search confidence level
({\bf scl}). The second criterion rejects candidates from the
preliminary list using statistics of Voronoi Tessellation for a
poissonian distribution of particles \citep{kia96}, by computing the
probability that an overdensity is a random fluctuation.  This is
called the rejection confidence level ({\bf rcl}).  More details can
be found in \citet{ram01}.

\citet{kim02} used a color-magnitude relation to divide the galaxy
catalog into separate redshift bins, and ran the VT code on each
bin. The candidates originating in different bins were
cross-correlated to filter out significant overlaps and produce the
final catalog. \citet{ram01} follow a different approach, as they do
not have color information. Instead, they use the object magnitudes
to minimize background/foreground contamination and enhance the
cluster contrast, as follows:

\begin{description}
\item[(i)] The galaxy catalog is divided into magnitude bins,
starting at the bright limit of the sample and shifting to progressively
fainter magnitudes. The step size adopted is derived from
the photometric errors of the catalog.

\item[(ii)] The VT code is run using the galaxy catalog for each bin,
resulting in a catalog of cluster candidates associated with each
magnitude slice.

\item[(iii)] The centroid of a cluster candidate detected in different bins
will change due to the statistical noise of the foreground/background
galaxy distribution. Thus, the cluster catalogs from all bins are
cross-matched, and overdensities are merged according
to a set criterion (described below), producing a combined catalog.

\item[(iv)] A minimum number (N$_{min}$) of detections in multiple
bins is required in order to consider a given fluctuation as a cluster
candidate. N$_{min}$ acts as a final threshold for the whole
procedure.  After this step, the final cluster catalog is complete.
\end{description}

\citet{ram01} applied their algorithm to one of the Palomar Distant
Cluster Survey (PDCS) galaxy catalogs.  They divided this deep data
into bins of two magnitudes, starting with $18.0 \le V_4 \le 20.0$ and
shifting to fainter bins in steps of 0.1 mag down to the detection
limit ($V_4 = 23.8$). This procedure resulted in 39 bins.  They adopted
{\bf scl} = 0.80 and {\bf rcl} = 0.05 to select fluctuations in each
bin. They then merged candidates whose centers had a projected
distance equivalent to $d_{12} \le 0.3min(R_1,R_2)$, where $R_1$ and
$R_2$ are the radii of any two candidates being compared.  Finally,
they kept only candidates with N$_{min} = 5$. While not stated in
their paper, they further merged clusters whose radii overlap.

When applying this algorithm to DPOSS data we tested different bins
and step sizes. As with the PDCS galaxy catalog, we span five
magnitudes, but in a brighter regime and with larger photometric
errors. We adopt a bin size of 1.5 magnitude and a step size of 0.2
magnitudes.  This step size is comparable to the photometric error at
$m_r = 19.5$. A wider magnitude bin could be used, but would
significantly decrease the number of steps.  A narrower bin would
result in a low number of cluster galaxy counts per bin.  We thus have
19 bins spanning the range $16 \le m_r \le 21.1$.  As seen in Figure
2, 1.5 magnitudes is a reasonable range to see clusters over the
background.

The threshold given by {\bf scl} selects overdensities above a given
galaxy surface density. Due to the non-uniform nature of plate data
and the effects of large scale structure, we do not use a single
threshold for the 108 different plates used in this project. Instead,
we maintain {\bf rcl} fixed at 5\%, but allow {\bf scl} to vary. The
VT code is run for each DPOSS field varying {\bf scl} from 0.78 to
0.92 (in steps of 0.02), instead of adopting a fixed value of 0.80 as
in \citet{ram01}.

The percolation analysis we apply to the cluster candidate catalogs
from different bins is similar to that of \citet{ram01}. However, the
optimal parameters found for DPOSS data are slightly different.  We
chose $d_{12} \le 0.8min(R_1,R_2)$ and N$_{min} \ge 3$, plus a final
merging of structures whose separation is smaller than the radius of
the largest neighboring candidate. These values are selected based on
the redshift range over which we are detecting clusters, and the
nature of our data. The PDCS data used by \citet{ram01} has no nearby
large clusters, as the data covers only 1 $\square ^\circ$ and was
designed to avoid bright objects and low-$z$ clusters. Tests done with
a smaller matching radius show that many nearby clusters would be
broken into subcomponents. The use of larger radii can incorrectly associate
adjacent clusters into a single candidate. The choice of N$_{min} \ge
3$ is explained in the next section, as it depends on the number of
false clusters (the contamination rate) produced for each DPOSS field. The
choice is sensible given that we have half the bins (19 compared to 39)
of \citet{ram01}.

As with the AK, the VT is optimized based on simulations
of the background galaxy distribution, where we apply the VT code
to both these simulated fields and real DPOSS fields. These techniques are
described in \S4.

\subsection{The Adaptive Kernel Technique}

We use a two-stage version of the adaptive kernel technique, which generates
an initial density estimate on a fixed grid. It then applies a
smoothing kernel whose size changes as a function of the local
density, with smaller kernels at higher densities. Finally, a 
density map is generated for each DPOSS field and SExtractor is used to locate
overdensities on this map. More details can be found in Papers I and II
and in \citet{sil86}. The galaxy catalogs used as input to the AK
have galaxies at 19 $\le m_r \le$ 21.1. In the previous NoSOCS papers 
the pixel scale adopted for the density maps was 1 pixel = 60$''$. For
the current project, we aim to detect distant (generally compact, low 
contrast) structures. We therefore require higher resolution for
the density maps, which are here generated with a scale of 1 pixel = 10$''$.

\citet{gla00} utilize a more powerful technique to detect clusters,
with a color-magnitude relation as a filter to minimize contamination
from foreground sources. However, the surface density of objects is
evaluated in a similar way to ours. The main difference is that they
use a simple fixed kernel, which is applied independently to 
each color slice. They argue that our estimator might not be
optimal because the AK tends to resolve high density regions in the
data. Based on cluster simulations, we show (see Figure 1 in Paper II)
that with the proper choice of the initial kernel size we do not divide
nearby rich clusters into subcomponents, while remaining sensitive
to low contrast structures (poor and close, or rich and distant). A
simple fixed kernel technique may be inappropriate in our situation, as
it is not simultaneously sensitive to clusters of different richness
classes at different distances. 

In Papers II and III we adopted an initial 
kernel of $500''$ radius to look for
lower redshift clusters ($z < 0.3$) with our brighter galaxy
catalogs. As before, we use simulated clusters to test the kernel size
for the fainter catalogs at $19 \le m_r \le 21.1$. The optimal value
found is $260''$.  Figure 9 shows a density map generated with a
$260''$ radius initial kernel. 48 artificial clusters of six richness
classes (N$_{gals} = 15, 25, 35, 55, 80, 120$) at 8 different
redshifts ($z = $0.15, 0.20, 0.25, 0.30, 0.35, 0.40, 0.45, 0.50) are
inserted.  Clusters are marked with circles and each column represents
a different richness class (increasing from left to right), while each
row is a different redshift (increasing from bottom to top). We can
see that our choice for the initial kernel radius does not break up
the nearby clusters, while it is still sensitive to distant rich
structures.

After generating a density map for each DPOSS field we then run SExtractor
to detect high density peaks, which we associate with cluster candidates.
Simulated background distributions are used to optimize SExtractor
parameters (see section 4). As in Papers II and III, each DPOSS 
field is optimized independently.

\section{Contamination and Optimization of the Algorithms}

As stated in Paper II, contamination by random fluctuations
in the galaxy distribution and chance alignments of galaxy groups
can seriously affect any cluster catalog. This problem is exacerbated
in a survey like this one, which lacks color information to minimize
background/foreground contamination \citep{gla00, got02}.

The most common methods to estimate how much a given catalog is
affected by these factors use simulated background distributions.
A simple random distribution with the same number density as the
original field would underestimate contamination. A background
constructed this way reproduces random fluctuations, but not the large
scale structure present in a given field. \citet{got02} tested three
different backgrounds, where they: (i) shuffled the positions without
touching colors; (ii) shuffled the colors, but did not touch the
positions; (iii) shuffled the colors and smeared the positions. That
was done specifically for their algorithm, which makes use of the
extensive multicolor information from the SDSS. Algorithms which do
not make use of color usually have their contamination estimate based
on a shuffled background \citep{lob00}, a Poissonian galaxy
distribution \citep{ram01}, or on a distribution generated with the same
angular correlation properties as the real field (P02; Paper II;
Gilbank 2001).

For this project we have tested four types of simulated backgrounds:
(i) a Raleigh-Levi (RL) distribution (P02; Paper II); (ii) a shuffled
background; (iii) a randomized galaxy distribution; and (iv) a smeared
background. In the latter case galaxy magnitudes are maintained, while
the positions are randomly redistributed within 7$'$ of their original
ones. \citet{got02} adopted 5$'$ for the smearing scale. They found
the contamination rate evaluated with this background to be extremely
high in comparison to the other ones tested.  As described below, we
test the optimization of our algorithms at the 5\% and 10\%
contamination levels. Commensurate with the results of \citet{got02},
we find that it is not possible to optimize the AK or VT at these
levels with the smeared background, and we have therefore abandoned it
in the tests described below.  The RL distribution is intended to
simulate galaxy distributions with an angular two point correlation
function similar to that of the real data.  

A variety of methods have been used to minimize the number of false detections
(N$_{false}$) in a simulated background 
distribution when optimizing cluster detection 
algorithms. The number of false clusters per square degree has been 
presented as an indicator of how low
contamination effects could be for a given catalog. However, as 
the photometric depth of catalogs varies, so will the absolute number 
of false detections. Thus, the contamination rate
($C$ = N$_{false}$/N$_{real}$) has been more commonly used to
optimize cluster detection algorithms. N$_{false}$ gives the number
of false detections in a simulated distribution, while N$_{real}$
is the number of detections in the real galaxy distribution. Optimal
detection thresholds are achieved for the highest recovery rate allowed
at a given contamination level.

We performed tests using 20 DPOSS fields and compared the results
obtained at the 5\% and 10\% contamination levels.  For the VT algorithm
the threshold is given by a combination of {\bf scl} and
N$_{min}$ (see section 3.1). We chose not to
fix the values of these parameters; rather, they are determined
independently for each field. For a given DPOSS field the best option might
consist of a high value of {\bf scl} with a low number of minimum
detections, while the reverse could be true in other cases.

Figure 10 compares the variation of N$_{false}$ with N$_{min}$ 
(left panels) and {\bf scl} (right panels) for the
RL, randomized and shuffled distributions.  Each dashed line
represents the variation of N$_{false}$ for one of the 20 DPOSS plates
tested. The solid line shows the mean variation on each panel. 
In order to compare the
performance of the three different backgrounds we use N$_{false}$, as
it is the only parameter to vary when estimating $C$.  
The dispersion is basically the same for
all distributions in the left panels, being slightly higher 
when the shuffled background
is used. The right panels indicate the contamination estimate is lowest, and
least sensitive, to the randomized background. The RL and shuffled
distributions show similar results to each other (except for two
outliers in the latter case). This suggests that the randomized
background underestimates the contamination. The RL is a model
representation of the angular distribution in a galaxy
field. N$_{false}$ smoothly decreases as we move to higher thresholds
(given by N$_{min}$ and {\bf scl}). However, the RL distribution might
not properly represent variations in high density fields. We expect
the correlation properties to vary for different DPOSS fields as they
have different galaxy densities \citep{mad96}. 
As a full analysis
of the correlation properties of all DPOSS fields is beyond the scope
of this project, we decided not to use the RL distribution for this
paper.  Instead, the shuffled background is adopted to optimize our
algorithms.  It shows no large differences relative to the RL
distribution, and is a model-independent representation of the
background.

We then generate a shuffled catalog (simply repositioning
all galaxies within a DPOSS field and keeping the magnitudes)
and run the VT on this list, which gives
N$_{false}$. The VT is also run on the real catalog, yielding
N$_{real}$. Each field is optimized for the values of {\bf scl} and
N$_{min}$ which yield the maximum N$_{real}$, while
keeping the contamination rate fixed at $C\sim5\%$. Note that 5\%
is a lower limit for contamination effects. A full assessment requires
a spectroscopic follow-up. 

The strategy adopted to optimize the AK algorithm is basically
the same as for the VT. We take the shuffled catalogs generated
for the VT optimization and trim them to the appropriate magnitude limit
($19 \le m_r \le 21.1$) adopted for the AK. We run the AK on these catalogs
as well as the original data to generate density maps for each.
SExtractor is used to detect overdensities on these maps.

The parameters that give the final optimization of the AK are the
minimum area and the threshold required for a detection when running
SExtractor. The minimum area gives the minimum number of pixels a
candidate must have, while the threshold gives the minimum number of
galaxies per square degree.  We vary the threshold from 3000 to 9000
galaxies per $\square^\circ$.  Tests in a broader range for 20 plates
show that the optimal threshold is always found within these values.
Four different values are tested for the minimum area: 200, 300, 500
and 900 pixels$^2$. In order to choose the optimal value for the
minimum area we evaluate the selection function with these four
different values (see Figure 11). We find that 200 and 300 pixels$^2$
produce similar results, with a slightly higher recovery rate in the
former case.  Going to 100 pixels$^2$ does not result in any
improvement. Thus, we adopted 200 pixels$^2$ as the minimum area for
all DPOSS fields when running SExtractor. As stated above the threshold is
optimized for each plate.

\section{The Cluster Catalogs}

The main goal of this project is to provide a catalog with rich galaxy
clusters to intermediate redshift for follow-up studies.
In order to be able
to select subsamples from the main catalog we provide some
basic properties, such as redshift and richness, 
for the cluster candidates. When
estimating these two quantities we use similar, 
but not identical, techniques to those
employed in Papers II and III. However, we face a variety of challenges
for the current project.
For instance, the photometric errors are large in the magnitude 
range utilized here, which also affects color measures; the 
number of clusters with known spectroscopic redshift
to be used as a training sample is small when we go out to 
$z \sim 0.5$; the $(g-r)$ color is not useful for $z \gtrsim 0.4$ 
clusters (as the 4000 {\AA} break shifts from $g-r$ to $r-i$); and
the blue plates do not necessarily go as deep as the red ones.

In \S5.1 and \S5.2 we describe our efforts to estimate redshift and richness
for the cluster candidates. A visual inspection is employed in \S5.3 
to eliminate obvious false clusters associated with bright stars, 
nearby galaxies and groups. A combined version of the VT and AK cluster 
catalogs is presented in section 5.4.

\subsection{Photometric Redshifts}

An empirical relation based on the mean $r$ magnitude and median $(g-r)$
color was successfully used to estimate redshifts for the NoSOCS clusters
(Papers I, II and III). Here, we assume these same properties are strong
indicators of the redshift of a cluster. However, this paper employs
significantly different methodologies, and it is not clear {\it a priori} if
useful photometric redshift estimation is possible using the 
faintest DPOSS data.

The main differences with respect to the photometric redshift technique
employed in Paper II are: (i) The galaxy catalogs
used here have a statistical correction applied to them (section 2.2), which
gives rise to a $\sim 15\%$ difference in number counts around $m_r = 20$;
(ii) counts were previously done at 15 $ < m_r < $ 20, while now we
count galaxies at 16 $< m_r < 21.1$; (iii) we now adopt local
estimates for the background counts, instead of taking
these estimates from the whole area of each plate. Additionally,
we tested different counting radii (0.50, 0.75, 1.0 and 1.5 h$^{-1}$ Mpc),
and found that 1.0 h$^{-1}$ Mpc produces the fewest outliers
and minimizes the overall dispersion.

The first step towards the determination of photometric redshifts is
the compilation of a list of clusters with measured spectroscopic
redshifts.  Unfortunately, there are few clusters with spectra
taken at $z > 0.3$, which biases our sample to low-$z$ clusters. Our
calibration sample consists of 238 clusters over the $\sim 2,700$
square degrees of this survey, taken from \citet{str99},
\citet{hol99}, \citet{vil98}, \citet{cal96}, and \citet{mul03}. In
comparison, the training sample for Paper II had 369 clusters, over a
narrower redshift range.

We proceed as follows: each of the 238 clusters with known
spectroscopic redshift has the number of galaxies as a function of
magnitude (N$_r$) and color (N$_{(g-r)}$) determined. These counts
have the local background counts subtracted, resulting in a
determination of the net cluster counts as a function of both color
and magnitude. The mean $r$ magnitude ($r_{mean}$) and median color
($(g-r)_{median}$) is then computed for each cluster. Then we bin the
colors and magnitudes in redshift bins of $\Delta z = 0.05$,
calculating the mean $z_{spec}, r_{mean}$ and $(g-r)_{median}$ for
each bin (see Figure 12).  These values are used to derive two
empirical relations for redshift estimation.  The first uses only
$r_{mean}$, while the second uses both $r_{mean}$ and
$(g-r)_{median}$.

In Figure 12 we show the dependence of the estimated values of
$r_{mean}$ and $(g-r)_{median}$ on spectroscopic redshift.  As
expected, colors are well behaved to $z \sim 0.4$. At higher redshifts
we expect the $(g-r)$ color to remain approximately constant, as the
4000\AA~ break shifts from the $(g-r)$ color to $(r-i)$. However, we
see that the $(g-r)$ estimate becomes meaningless for $z \gtrsim 0.4$.
Two factors contribute to this color mismeasurement. First, there is
incompleteness in the J (blue) catalogs. As seen in \S 2.2, the field
selection is based only on the F (red) plates.  We select only those F
plates which go as deep as $m_r = 21$, but the corresponding blue
plates might not be equally deep.  Fainter than $m_r \sim 19.5$
the number of two-band detections drops steeply, reaching
$\sim 70\%$ at $m_r = 21$. For Papers I, II and III the magnitude limit
of the galaxy catalogs ($m_r = 19.5$) allowed us
to require a $g$-band detection when generating a galaxy catalog for
the cluster search. This is not possible for the current survey. The
second factor contributing to color mismeasurement is the possible
contamination from stars at faint magnitudes.

In Figure 13 we compare the photometric and spectroscopic redshifts
from the two relations discussed above, using only magnitude (left
panel) and both color and magnitude (right panel).  We find the
dispersions to be similar, being slightly higher when we include
colors to evaluate $z_{phot}$. As the color information does not
significantly improve the photometric redshift estimate (using our
data), we estimate redshifts using only $r_{mean}$. The fitted
relation is
$$ z_{phot} = 8.93 - 1.10 \times r_{mean} + 0.03 \times r^2_{mean}
\eqno(2) $$ 
with a dispersion in redshift of $\Delta z = 0.052$. These
errors are comparable to those of other techniques which rely solely on
magnitudes for redshift estimation, such as the matched filter.

If we have no a priori knowledge of the cluster redshift, it must be
determined iteratively. We choose an initial guess of the
redshift ($z_{start} = 0.15$), and compute $r_{mean}$ and $(g-r)_{median}$
based on the cluster corrected counts within 1.0 h$^{-1}$ Mpc,
apply the empirical relation derived above and derive a photometric
estimate of $z$. We then iterate this procedure until it converges,
adjusting the radius in each iteration.

Using the 238 clusters with known $z_{spec}$, we tested four different
starting redshifts ($z_{start} = 0.05, 0.10, 0.15, 0.20$). 
We find the estimates to have almost no dependence
on the initial redshift. For instance, when comparing the results with
$z_{start} = 0.05$ and $z_{start} = 0.15$, we find that
$Q_{sigma}$ is $(z_{0.15} - z_{0.05})/(1 +
z_{0.15}) = 0.005$.  
We chose to use $z_{start} = 0.15$, as it results in fewer outliers.

Figure 14 shows the redshift distribution for different richness classes
using both the AK and VT. The redshift cutoff in each case is generally
in agreement with that seen in the selection functions (\S6).

\subsection{Richness Estimates}

Richness is evaluated in a similar manner to Papers II and III.
We describe below all the steps
in the richness determination, pointing out
the minor differences from the previously employed procedure.
We proceed as follows:

\begin{enumerate}
\item We count the number of galaxies at $16.0 \le m_r \le 21.0$ within
1.0 h$^{-1}$ Mpc of the cluster center. The background counts in
the same range are evaluated locally, scaled to the cluster area, and
subtracted, yielding the background-corrected cluster counts
(hereafter N$_{corr}$). In the previous papers the magnitude range
was $15.0 \le m_r \le 20.0$ and the background was estimated from the
whole plate.

\item We run a bootstrap procedure with 100 iterations. In each iteration, we
randomly select N$_{corr}$ galaxies at $16.0 \le m_r \le 21.0$, within
1.0 h$^{-1}$ Mpc.
Each galaxy has its apparent magnitude converted to an absolute
magnitude. As with the artificial clusters (\S6), we consider the clusters
to be composed of 60\% early-type and 40\% late-type galaxies. When
transforming to absolute magnitudes, we simply use elliptical K-corrections
for 60\% of the objects, and Sbc K-corrections for the remainder. We then
count the number of galaxies between M$^*_r$-1 and M$^*_r$+2, where
M$^*_r$ = -21.52 \citep{pao01}. The mean value from
the 100 iterations is computed, yielding the cluster richness N$_{gals}$,
which can be decomposed into 40\% Sbc galaxies (N$_{gals,Sbc}$) and
60\%  ellipticals (N$_{gals,E}$). If the cluster has
a redshift such that M$^*$-1 to M$^*$+2 is fully sampled in the
range $16.0 \le m_r \le 21.0$, then the procedure is finished here.
Otherwise we have to apply a correction factor to the richness
estimate (N$_{gals}$), as described in step 3.

\item If the cluster is either too nearby or too distant, then M$^*$-1
$<$ M$_{16}$ or M$^*$+2 $>$ M$_{21}$, respectively, where M$_{16}$ and
M$_{21}$ are the bright and faint absolute magnitude limits
corresponding to $m_r = 16.0$ and $m_r = 21.0$. In practice these
limits are different if a given galaxy is considered to be elliptical
or spiral, due to the differing K-correction factors. Whenever M$^*$-1
to M$^*$+2 does not lie within M$_{16}$ and M$_{21}$ we apply
correction factors for the richness estimate, defined as:
$$\gamma_1 = {\int_{M_r^*-1}^{M_r^*+2} \Phi(M)dM \over
\int_{M_{16}}^{M_r^*+2} \Phi(M)dM} \eqno (3)$$
$$\gamma_2 = {\int_{M_r^*-1}^{M_r^*+2} \Phi(M)dM \over
\int_{M_r^*-1}^{M_{21}} \Phi(M)dM} \eqno (4)$$ We call $\gamma_1$ and
$\gamma_2$ the low and high magnitude limit correction
factors. Actually, each of these correction factors are evaluated
twice, once for the magnitude limit for elliptical galaxies and the
other for late-types. Whenever necessary, one of the above factors (as
we span 5 magnitudes it is impossible to simultaneously miss both the
bright and faint end) is multiplied by N$_{gals,E}$ and/or
N$_{gals,Sbc}$.
\end{enumerate}

The main differences between steps 2 and 3 in this work relative to the
procedure adopted for the previous NoSOCS papers lie
in the assumption that the cluster is not totally composed
of elliptical galaxies, and that we consider
a correction factor to the lower magnitude limit.

In Figure 15 we show the richness distribution for the VT
and AK cluster candidates for different redshift bins. The
top panel shows the richness distribution for the entire sample.
Further considerations related to redshift and richness estimates are
discussed in section 8, where we compare the current catalog
to the previous NoSOCS cluster catalog and to other surveys.

\subsection{Elimination of Spurious Cluster Candidates}

Before combining the VT and AK cluster catalogs we 
perform a visual inspection of the cluster candidates to eliminate obvious 
false clusters. Similar procedures 
were previously adopted by \citet{kim01} and \citet{got02}. While
inspecting the candidates we realized that many of them were associated
with bright stars, and some with nearby bright galaxies and groups,
globular or open clusters, plate defects, trails, etc. Thus, we
decided to try to eliminate most of these false clusters in an automated
fashion.

The major source of spurious detections is bright objects which were
missed when generating the list of bad areas to be excised. In some
cases, the object was removed, but the excised area was not large
enough. We therefore decided to compare the cluster candidate
positions to a bright star catalog. We used the Tycho-2 catalog
\citep{hog00}, which is $\sim$ 90\% complete to $V \sim 11.5$.

Figure 16 shows, for 4 different magnitude bins, the distribution of
offsets (in arcsec) between Tycho-2 stars and the nearest cluster
candidate. The solid lines show the real offset distributions, while
the dotted lines show the offset distribution between a mock star
catalog and the cluster catalog.  At 8 $< V <$ 10, almost all stars
are associated with a cluster candidate within $120''$. Our
photometry fails for these bright stars, which gives rise to a large
number of faint, spurious detections in the halos of these
objects. These spurious objects generate overdensities detected as
cluster candidates. The catalogs presented in Papers I, II \& III did
not face this problem as most of these spurious detections are fainter
than $m_r = 19.5$, and do not have counterparts in the $g$-band.  For
the single band galaxy catalog utilized here, the elimination of
bright objects turns out to be a serious issue.

We use the above information to clean our cluster catalog in a
semi-automated fashion. From the inspection of offset distributions in
different magnitude bins, we selected a minimum offset between a star
and a cluster candidate (in each magnitude bin) to exclude the cluster
as a false detection, and a function is interpolated to these minimum
offsets as function of magnitude. $$ \Delta\theta_{min} = 185 + 0.37
\times (V^2) - 0.01 \times (V^4) \eqno(5) $$ This function gives the
minimum separation (in arcsec) a given cluster candidate must have
from a star of a given V magnitude to be retained in the catalog. If
the separation is smaller than this value, the candidate is
excluded. Adopting this procedure, we eliminated 1031 VT and 1303 AK
candidates. We further compared the cluster candidates to lists of
bright galaxies, and globular and open clusters, which resulted in the
elimination of a few more candidates. Finally, we visually inspect all
the remaining candidates to exclude any associated with plate defects or
trails. The total number of eliminated cluster candidates is 1602
($\sim 17\%$) and 1748 ($\sim 27\%$) for the VT and AK catalogs,
respectively. Note that as the detection codes were optimized to
produce 5\% contamination, reducing the number of real candidates by
20\% results in an increase of the contamination rate to $C$ =
N$_{false}$/N$_{real} \sim 6\%$.
  
\subsection{The Northern Sky Optical Cluster Survey (NoSOCS) 
High-Redshift Catalog}

The NoSOCS supplemental catalog presented here covers $\sim$ 2,708
$\square^{\circ}$ and originates from two cluster detection
algorithms. The catalog contains 9,956 cluster candidates, which
translates into a surface density of $\sim$ 3.7 clusters per square
degree. The median estimated redshift of the AK clusters is 0.32, and
is 0.24 for the VT clusters. The median richness is N$_{gals}$ = 51.2
for the AK and N$_{gals}$ = 29.3 for the VT.  The sky distribution for
the combined AK-VT cluster candidates is shown in Figure 17, for the
regions in the northern galactic hemisphere (NGH) and southern
galactic hemisphere (SGH).  Examples of rich and distant clusters
detected by the VT and AK codes are shown in Figure 18. The images are
taken from the DPOSS F plate and span 250$''$ on a side. However, the
low quality of the plates makes it difficult to visually identify most
of these distant clusters.  Figure 19 compares a cluster candidate
imaged with a CCD at the Palomar 60$''$ telescope with its
corresponding DPOSS F plate image.

The combined cluster catalog with 9,956 candidates is presented in
Table 1, which is sorted by RA. In this printed version we present a
sample of clusters with $0.4 \le z_{est} \le 0.5$ and N$_{gals} \ge
60.0$ (using the AK estimates).  A complete version of this table can
be found online at \url{http://dposs.ncsa.uiuc.edu/} and the
associated mirror sites. The table is organized as follows: column 1
gives the cluster name, where the convention is NSCS Jhhmmss+ddmmss
(the second ``S'' in the name stands for {\it supplemental}); columns
2 and 3 give the RA and Dec in decimal degrees (J2000); the redshift
estimates are shown in columns 4 (AK) and 5 (VT).  Entries are set to
zero whenever the corresponding estimator failed or if the candidate
was not detected by that algorithm. Columns 6 and 7 present the
richness estimates.  The few candidates with redshift estimates
greater than 0.6 (less than 40 clusters for each algorithm) are marked
with the string ``---'' and have their richness set to zero.  A few
other candidates with the N$_{gals}$ estimate greater than 400 (less
than 10 clusters on each case) have their richness set to zero.
We decided to keep redshift estimates at $0.5 \le z_{est} \le 0.6$
instead of setting high-z estimates to a given value, such as $z = 0.5$
(as done by \citet{kim01}), and accumulate candidates in the 
last redshift bin.
Column 8 has the plate number, while the codes that retrieved the
cluster are listed in column 9. This column has ``A'' for the AK code,
``V'' for the VT and ``AV'' for both. We also provide (electronically) 
lists of bad areas used, as well as some other useful
information for the reader. Cluster candidates found within 500$''$ of the
plate edges (nearly two times the size of the initial kernel for the AK)
are marked with an ``*'' in the last column. There are 333 candidates
in these regions ($\sim$ 3\% of the total sample); they are marked
because their proximity to a plate edge may affect the 
local density estimate.

\section{The Selection Function}

In order to estimate the completeness of this catalog, we evaluated the
selection function (SF) for a small number of DPOSS fields.  Unlike Papers
II and III, we do not perform a complete field-by-field analysis of
the SF. However, it is important to note that 
for statistical studies such as the estimation of the space density
and mass function of clusters, the SF should be evaluated individually 
for each field.
The SF measurements  presented here provide an approximate
completeness estimate, and are used to check for possible improvement
in the recovery rate for rich clusters when allowing a higher
contamination level. In addition, the SF is useful for determining if
the estimated redshift distributions shown in Figure 14 are
realistic. Finally, in the richness and redshift regime where we
predict our catalog to be nearly complete, we expect to find good
agreement with other catalogs of rich clusters spanning the same
volume. Comparisons to such catalogs are presented in \S 8.

The most common technique to evaluate the selection
function makes use of artificial clusters.
These are inserted in a background distribution
and their recovery rate as a function of richness and redshift
gives the SF. However, there are three major complicating factors:

\begin{enumerate}
\item The background used for completeness tests should not be the same
as the one used for the optimization of the algorithm. As shown
by \citet{got02} and in Paper II, a simulated background
overestimates the completeness rate. The real background distribution should
be used instead.

\item When randomly placing artificial clusters in the background
distribution we must avoid bad areas (due to saturated objects), as well as
the positions of cluster candidates, as done by P02 and in
Paper II. A simulated cluster located in the vicinity of
 a real cluster has its recovery probability changed from what
it would be in a region devoid of clusters (which is more likely to be
a background region). In this way we avoid the bias in the detection
rates discussed by \citet{got02}.

\item The background plus artificial clusters galaxy catalog
must have the same number density as the real field, as the
optimal threshold was obtained with the original number density.
Galaxies from rich artificial clusters can increase
the number density of the field by a few percent.
\end{enumerate}

We generate sets of 48 simulated clusters with
6 different richness classes (N$_{gals} = 15, 25, 35, 55, 80, 120$)
and placed at 8 different redshifts ($z = 0.15, 0.20, 0.25, 0.30,
0.35, 0.40, 0.45, 0.50$). These clusters are placed at random 
locations within a field; if it falls on a bad area or cluster candidate, 
a new position is chosen.  We also
preserve the same number of objects as in
the original field by randomly removing the same number of galaxies
from the original catalog as are inserted with the artificial clusters.
This procedure is repeated 50 times.

For both the AK and VT methods we use a 0.75 h$^{-1}$ Mpc radius when
comparing the input and output positions (this choice is explained below).
An example of the selection function obtained for a
DPOSS field with the VT code is shown in the middle panel of Figure 20.
The SF for the AK code with the same field is shown in the bottom panel of 
Figure 11.

Artificial clusters were generated in the same way as in
Paper II. They follow a Schecter luminosity function, with
parameters given by \citet{pao01}. The characteristic
magnitude is $M^* = -21.52$, while $\alpha = -1.1$. Cluster galaxies
lie at $-23.4 \le M_r \le -16.4$ and are composed of 60\%
elliptical galaxies and 40\% Sbc galaxies. K-corrections are obtained
through the convolution of SEDs taken from Coleman, Wu \& Weedman (1980),
with the DPOSS $r$ filter.  We adopt a power law in radius for
the surface profile, $r^\beta$, where $\beta=-1.3$ lies in the middle of
the observed range \citep{squ96, tys95}.
Clusters also have a cut-off radius at $r_{max} = 1.5$ h$^{-1}$ Mpc and
a core radius $r_{core} = 0.15$ h$^{-1}$ Mpc.

In Paper II all of these five basic parameters (luminosity profile slope,
$r_{max}$, $r_{core}$, spatial profile slope ($\beta$) and the cluster
composition) had their effects on the SF tested.
The strongest variation is with the choice of $\beta$. Large fluctuations
in the SF were not seen when varying the other parameters. We then
decided to adopt the canonical values above. In this work, 
significant differences could arise from testing different cluster 
compositions. At higher redshifts, the K-correction effects could be more
significant when testing clusters with different compositions.
Figure 20 shows a comparison of the SF for the VT code evaluated with three
different cluster compositions (20\% E \& 80\% Sbc; 60\% E \& 40\% Sbc;
100\% E). There is a clear trend toward a higher recovery rate
when using clusters mainly composed of late-type galaxies,
although the effect is weak. If we fix the comparison at $z = 0.35$
and N$_{gals} = 80$, the recovery rate decreases from
88\% to 84\% from the top to the middle panel,
and finally to 80\% for the bottom panel.

We also test for improvement in the SF when we change the required value of
$C$ from 5\% to 10\%. 
As there is no significant improvement for the most
distant and rich clusters when allowing higher contamination,
we keep the contamination rate fixed at $C = 5\%$.

When assessing the SF of a given cluster catalog, it 
is important to consider items (1), (2) and (3) listed
above, as well as the matching radius used to compare input
and output positions of artificial clusters.
\citet{got02} checked the dependence of the positional deviation
according to redshift and richness. They used a small radius
of 1.2 arcmin to match input and output catalogs, and found a negligible 
variation to $z \sim 0.4$.
Instead of using an apparent radius for the match, we use
a physical radius. We perform a test with 1.5 $h^{-1}$ Mpc (the size
of the input clusters), checking the dependence of
the offsets on redshift and richness. 

Figure 21 shows (for both the AK \& VT) the positional offset (in Mpc)
according to redshift for different richness classes.
Each line represents a different richness class. Thicker lines
represent richer clusters. For the poorer clusters the offset goes to
infinity for higher redshifts ($z> 0.40$). This simply means the
recovery rate goes to zero for these clusters.
We note that the offset is slightly larger and noisier for the VT code as
compared to the AK. This is due to the percolation analysis needed for
the VT, where we begin with candidates in different magnitude bins.
An artificial cluster can have its recovered position slightly changed
if it percolates with a nearby fluctuation.

We have also tried to evaluate the SF with four different
matching radii for the VT code: 0.4, 0.75, 1.0 and 1.5 $h^{-1}$ Mpc.
There is a great improvement when we go from 0.4 $h^{-1}$ Mpc
to 0.75 $h^{-1}$ Mpc, while the SF does not change significantly for
larger matching radii. Our final choice is
0.75 $h^{-1}$ Mpc for comparing the original positions
of artificial clusters with the output of both our algorithms.

Finally, it is important to stress that the simulations used to assess
the contamination rate (\S 4) only provide a lower limit. Similarly,
the cluster simulations employed provide only an upper limit for the
completeness rate of the catalog.

\section{Performance comparison of the two algorithms}

The AK cluster catalog contains 4,813 candidates, whereas the VT
contains 7,962. The combined catalog shown in Figure 17 contains 9,956
cluster candidates. Based on Figures 11 and 20, we expect the AK catalog
to have more rich clusters, and to go deeper, compared to the VT
catalog.  On the other hand, the VT catalog is more complete in the
regime of poor/nearby systems. The redshift and richness distributions
of the cluster candidates confirm the expectations from the
simulations. The cutoffs in every panel of Figures 14 and 15 show that
the cluster distributions are in good agreement with the SFs.

We stress that both methods applied here are simple density
estimators, which detect fluctuations in the projected galaxy
distribution above a given threshold. Both methods make no assumptions
whatsoever on any physical properties of galaxy clusters (e.g., the
luminosity function, colors of galaxy members, spatial profile,
etc). The AK code is sensitive to an initial smoothing scale, while
the VT does not distribute the data in bins. However, with the proper
choice of the initial kernel size, we expect the AK to be as sensitive
to irregular structures as the VT. We believe the differences shown
here arise from the way we minimize the background in each case. The
VT is applied to galaxies spanning a wide range of magnitudes (16.0
$\le m_r \le 21.1$), while the AK code is applied only to the faint
data of DPOSS (the bright cut is at $m_r = 19.0$). Thus, the AK
naturally avoids nearby fluctuations and enhances the contrast of
distant systems, while we employ a binning scheme to minimize
contamination for the VT algorithm.  Obviously, these same density
estimators when applied to multi-wavelength data provide a much more
powerful tool to minimize background/foreground contamination
\citep{kim02, got02, gla00}.

The comparison of catalogs generated by these different techniques
is not a straightforward task.
As noted by \citet{kim01}, clusters are missed by one
method and not by another due to the fact that they are {\it seen} with
different {\it eyes}. Some cluster finding techniques will be biased
to clusters with some specific properties, thus having a
low efficiency for overdensities that do not match these
properties. Furthermore, clusters which are recovered by various
techniques might have large differences in the measurement of
properties such as richness, redshift and projected density profile.

\citet{kim01} used a 0.7 h$^{-1}$ Mpc radius when matching candidates from
different algorithms. However, she finds that, in some cases, the 
cluster centers from different algorithms can differ by as much as 
the extent of the cluster. This could be especially true for poor and 
irregular clusters. Large differences
in the redshift estimates for candidates from the two codes would make
this comparison more difficult, or even impossible. However, we do
not expect the photometric redshifts to have a large variation in our case.
So, we decided to adopt a physical radius, instead of an apparent
radius when comparing the AK and VT catalogs. We use a 0.75 h$^{-1}$ Mpc
maximum radius, as done for the evaluation of the selection functions.

Out of the 4,813 (AK) and 7,962 (VT) cluster candidates, 2819 (59\% of AK and
35\% of the VT) are common sources. The VT catalog is denser than
the AK catalog because it is more complete in the regime of poor systems,
which are more abundant.
Figure 22 shows the offset distribution for the matched clusters
in Mpc (solid line) and arcseconds (dotted line). Most clusters
show no large offsets (less than 100$''$ or 0.40 h$^{-1}$ Mpc).
The combined catalog is presented in Table 1 and Figure 17.

Redshift and richness estimates are obtained post-detection.  They are,
however, sensitive to the cluster center, as the corrected cluster
galaxy counts will depend on this choice.  We then divide the common
clusters into bins of 40$''$ in centroid offsets, from 0$''$ to
160$''$. The residuals as a function of redshift for different offsets
are shown in Figure 23, with a clear trend to large residuals for
large offsets. Figure 24 shows the same effect, but for richness.
Finally, we investigate the dependence of the richness measurement on
the adopted redshift. Figure 25 is similar to the previous one, but
the bins are in redshift residuals (from $\Delta z = 0.01$ to 0.04, in
steps of 0.01), instead of cluster center offsets.  The top left panel
shows the results when cutting the sample at offset $< 40''$ and
$\Delta z < 0.01$. There is an evident improvement when comparing this
panel to the bottom left panels of this Figure and Figure 24.

We show in Figure 26 richness (top) and redshift (bottom)
distributions for VT only detections (dashed-dotted line), AK only
clusters (dotted line) and common candidates (heavy solid line). This
plot fully confirms the expectations from the selection function
estimates. We expect the VT code to perform better for poor, nearby
clusters, while the AK should go deeper when detecting rich
systems. This is clearly seen in Figure 26, where most of the VT-only
detections are nearby poor systems, while the AK-only clusters are
mainly distant and rich. The two catalogs overlap at intermediate
richnesses and redshifts.

Figure 27 shows the distribution of N$_{gals}$ with respect to the
estimated redshift for both the AK and VT. A similar dependence was
previously found by \citet{kim01}. The richer systems are rarely found
at low redshifts where we probe less physical volume per unit redshift
interval.  On the other hand, the poor clusters are easier to detect at
low redshift, where they are still seen as high-contrast systems. One
surprising result is that only the VT code apparently finds relatively
poor systems (N$_{gals} \lesssim 60$) at $z > 0.45$. For these faint
systems the redshift and richness estimates are more difficult to
measure as the number of galaxies used for these estimates is
typically low. The VT code is also less accurate than the AK algorithm
in determining the centroid of a cluster (Figure 21).  For these low
contrast systems, this can have a great impact to the redshift and
richness estimates, and could result in underestimation of the
richness. Note also that \citet{kim01} found the same trend for the VT
candidates when using SDSS data.
 
Finally, we test the overlap of both catalogs in the regime where they
are expected to have high completeness.  The SF predicts that both
cluster catalogs are nearly complete for rich clusters (N$_{gals}
\gtrsim 65$) at $0.2 < z < 0.3$ (Figures 11 and 20).  To be
conservative we assume a completeness level $ > 90\%$, and select all
cluster candidates from both catalogs for these richness classes in
this redshift range. We find 87 VT and 108 AK candidates. Out of the 87 VT
candidates, 84 have a match in the AK catalog, which represents an overlap
of $\sim 78\%$, in agreement with the expectations from the SF.

\section{Comparison with other cluster surveys}

A comparison of the NoSOCS and the Abell cluster catalogs was
previously done in Papers II and III. We would like to compare the
supplemental NoSOCS catalog presented here to other intermediate
redshift cluster catalogs. The best sources of comparison are given by
the preliminary SDSS catalogs \citep{kim01, got02}, which are used as
a reference for the comparison shown in the end of this
section. Additionally, as the DPOSS data used differs from
that used in Papers II and III, we also perform comparisons with
those catalogs. We also investigate differences in the
redshift and richness measurements. As stated before, there are some
fundamental differences in the estimates presented here and in the
previous DPOSS papers.

From the $>$ 12,000 NoSOCS (Papers II and III) cluster candidates over
the entire high latitude ($|b| > 30^{\circ}$) northern sky, we find
4,211 candidates within the 2,700 square degrees sampled here.
As done before, we adopt a 0.75 h$^{-1}$ Mpc radius 
when comparing two catalogs.
We find 2,638 clusters common to both catalogs. Figure 28 shows
the estimated redshift distribution (bottom panel) of NoSOCS only
clusters (Papers II and III) as dotted lines, while the dashed lines
represents the distribution of clusters present only in the current
survey. The distribution of common clusters is represented by the
heavy solid line.  It is clear that most new candidates found here
have $z_{est} \gtrsim 0.2$.  In the top panel of the same Figure we
show the ratio of common clusters to each of the catalogs. The dotted
line represents the ratio to the catalog from Papers II and
III, while the dashed line is the ratio of common clusters to the
supplemental catalog presented here.  Out to $z_{est}\sim 0.2$,
approximately 60\% of NoSOCS clusters are common to the new
catalog, with almost all of the common detections at low-$z$.

It is important to stress that poor clusters at $z < 0.2$ detected in
Papers II \& III might not be easily recovered using the methods
presented here. The AK catalog is obviously biased against these
systems due to the bright magnitude cut ($m_r \sim 19$), while the VT
code might have problems recovering these nearby, low contrast
structures. The binning scheme adopted to enhance the cluster contrast
might not be powerful enough to enable recovery of these systems.
Note also that the matching radius employed when comparing the two
catalogs is crucial to the analysis. Poor systems might have large
differences in the centroid determination, especially if the centroid
is given by galaxies in different magnitude ranges. \citet{got02}
adopted a 6$'$ radius to compare their catalog to other SDSS cluster
catalogs. We preferred to adopt a physical radius of 0.75 h$^{-1}$ Mpc 
to compare any two catalogs, even considering possible discrepancies in
the photo-z estimates.
In the regime where they are expected to have high
completeness ($0.1 < z < 0.2$ and N$_{gals} \gtrsim 65$), 
the cluster catalog from Papers II
and III has 38 rich clusters. The VT catalog
finds 33 of those, while the AK catalog finds 27. The mean overlap is
$\sim$ 79\%. For the few missing clusters we pointed the richness
estimator code to the coordinates given in Papers II \& III, also
adopting the redshift estimate given there. We found these systems to
be poor clusters in the deeper galaxy catalog used here. In other
words, the contrast of these clusters decreases when seen in a wider
magnitude range.

In Figure 29 we plot the NoSOCS photometric redshift estimates
\citep{gal03} against those from this paper (left panel). If we assume
the estimates from Papers II \& III are more correct than those
presented here, then we have some indication that our redshifts
are overestimated.  In the right panel we compare the richness
estimates for nearby clusters ($z < 0.2$). As stated in section 5.2,
there are many differences between the richness estimates presented
here and in the previous DPOSS papers. First, the galaxy catalogs are
deeper and have a statistical correction applied to them, which could
affect the cluster contrast. Second, we now adopt a local background
correction, while the background counts were previously taken from a
plate scale (Papers II \& III). In the previous papers, the $\gamma$
correction factor considered all galaxies to be elliptical, while here
we consider only 60\% to be ellipticals.

Finally we compare our catalog to the SDSS cluster catalogs
presented in \citet{kim01} and \citet{got02}. Both catalogs
cover an equatorial strip of $\sim 350 \square^{\circ}$. However,
most of the common clusters (between SDSS and DPOSS)
are found in the RA $\lesssim 44^{\circ}$ region, as
we have only two plates in the North Galactic Pole
region which are also in this equatorial strip. The
coordinate limits used to trim both SDSS catalogs, as well
as ours, are $-2.1^{\circ} \le \alpha \le 43.75^{\circ}, 
-1.27^{\circ} \le \delta \le +1.27^{\circ}$. There are 863 
clusters from \citet{kim01}, 1288 from \citet{got02} and
471 AK-VT candidates in this region. The catalog from \citet{kim01}
is the combination of two catalogs, using a Hybrid Matched Filter 
and a Voronoi Tessellation technique, and is hereafter called HMF-VT.

In Figure 30 we show the estimated redshift distributions of clusters
detected only by DPOSS (dashed line), common clusters (heavy solid
line), and SDSS only clusters (dotted lines). In the top panel the
DPOSS cluster catalog is compared to the SDSS Cut \& Enhance (CE)
catalog \citep{got02}, while in the bottom panel the comparison is
made to the catalog of \citet{kim01}. 
In both panels the unmatched 
detections have redshift estimates from their own detection method,
while matched clusters have our redshift estimates.  The HMF-VT
catalog \citep{kim01} recovers many more nearby (probably poor)
clusters than DPOSS, while the CE catalog \citep{got02} goes deeper
than DPOSS.  This result also points to intrinsic differences in the
clusters sampled by the two SDSS catalogs.

In the top panels of Figure 31, our redshift measurements are compared
to those from both SDSS catalogs. The estimates of
\citet{kim01} have an {\it rms} comparable to ours, while
\citet{got02} measurements are much more precise ($\Delta z <
0.02$). Despite the offset, we find a better agreement when comparing
our estimates to those from \citet{kim01} than those from
\citet{got02}.  This might be due to the fact that Kim's estimates use
magnitude information (as we do), while the CE estimates are based
only on the $g-r$ color (which limits high-$z$ estimates due to the
shift of the 4000\AA~ break to the $r-i$ color).

When comparing richness measurements we re-measure N$_{gals}$ for the
common clusters, using the coordinates and redshift estimates provided
in the SDSS catalogs. This procedure minimizes the effects seen in
Figures 23 to 25. The richness measure given by \citet{kim01} is
called $\Lambda_{cl}$, which represents the total cluster luminosity
within 1.0 $h^{-1}$ Mpc in units of $L^*$. N$_{gals}$ also considers a
1 $h^{-1}$ Mpc counting radius, but we only sample galaxies at $M^*-1
\le M \le M^*-2$. \citet{got02} count galaxies from $m_3$ to $m_3 + 2$
(where $m_3$ is the third brightest cluster galaxy) within the
detection radius provided by the CE algorithm. As they state, this is
similar to the Abell richness, except for the counting radius, which
is chosen differently for the CE method.  The bottom left panel of
Figure 31 shows the relation between $\Lambda_{cl}$ and N$_{gals}$. As
we adopt the same coordinates and $z_{est}$ provided by \citet{kim01},
we expect the scatter to originate mainly from the differences in the
galaxy catalogs used for the richness estimation. Considering the
higher quality of the SDSS data compared to the DPOSS plate data, the
small scatter shown in this plot is an encouraging result. The bottom
right panel shows the comparison between the CE richness and
N$_{gals}$.  The scatter is extremely large and no obvious relation is
found. We believe this is unrelated to the nature of the data, but
mainly to the choice of an apparent radius to count galaxies for the
CE richness.  Even a direct comparison between the two SDSS
richnesses, $\Lambda_{cl}$ and that from the CE, is difficult.

The cluster number density presented here is over 2 times greater than that
found in Papers II and III.  This is sensible as we sample many more
distant clusters when compared to our previous papers.
The more interesting comparison, however, is to the SDSS catalogs.
The Gunn $r$-band has a similar response to the SDSS $r^*$ filter,
which renders our $m_r = 21.1$ limit similar to that adopted
by \citet{kim01} and a little brighter than the choice of \citet{got02}.
Thus, the cluster catalogs' depths should be similar. The
cluster catalogs will differ due to the quality of the
data used, the detection algorithms, and the contamination level allowed.

The number density of the HMF clusters is 7.8 clusters per square
degree \citep{kim01}, 13.3 for the CE catalog \citep{got02}, and 3.7
for our catalog. The main reasons that the HMF-VT catalog finds over
twice as many clusters as our technique are the higher quality
photometric data of SDSS and the higher contamination rate allowed by
\citet{kim01} (15\% compared to our 5\%). The CE catalog has an
extremely high number density. This is likely due to their fainter
magnitude limit, and their high ($\sim 30\%$) contamination rate in
the regime of poor clusters (where most candidates are). The CE method
is a powerful technique to detect clusters and minimize
background/foreground contamination, but the very
high number density of supposedly real clusters may not be justified.
A closer examination of our redshift distribution (Figure 14) when
compared to Figure 26 of \citet{got02} reveals that in the regime of
rich clusters our catalog goes as deep as all the SDSS
catalogs. However, the HMF-VT method samples many more poor/nearby
systems than our technique, while the CE catalog detects many more
high-redshift candidates ($z_{est} > 0.25$).  Nonetheless, it seems
that the SFs presented in \citet{got02} may not be in good agreement
with their redshift distribution. They appear to actually go deeper
than indicated by the SFs.

An illustration of the higher density of the SDSS HMF-VT catalog
compared to this AK-VT catalog is given in Figure 32.  We show the
candidate distribution for the central region of DPOSS field
824. HMF-VT candidates are shown as solid circles, while the AK-VT
candidates are plotted with dashed circles. The circles represent a
1.5 $h^{-1}$ Mpc radius at the estimated redshift. Two features are
readily noticed. First, the HMF-VT only clusters are generally low-$z$
systems as indicated by the large radii of the candidates missed in
our survey.  Second, the HMF-VT is able to recover cluster candidates
at different redshifts along the same line of sight, while our method
cannot distinguish between these superposed systems. In most such
cases, we recover the fainter, more distant clusters instead of the
closer ones.

Finally, as done in the comparison to the catalogs presented in Papers
II \& III, we test for disagreements with the SDSS catalogs where SDSS
and DPOSS predict high completeness levels. Using the relation between
$\Lambda_{cl}$ and N$_{gals}$, we convert the HMF-VT richness
estimates to N$_{gals}$. We look for clusters with N$_{gals} > 65$ at
$0.2 < z_{est} < 0.3$. There are only 2 clusters in this regime in the
HMF-VT catalog and 6 in our catalog. Both clusters found by
\citet{kim01} have a match in our catalog, while 3 of the 6 AK-VT
candidates have a match in the HMF-VT list. As we can not convert the
CE richness estimate to N$_{gals}$ we simply chose a richness cut for
the CE catalog. We selected all clusters with CE$_{rich} > 40$ at $0.2
< z_{est} < 0.3$. Out of 18 CE clusters in this regime, 11 have a
counterpart in our catalog. We would like to stress that 
centroid and redshift estimates discrepancies, as well
as the non-negligible scatter in the 
$\Lambda_{cl}-$N$_{gals}$ relation, render
this comparison problematic. Note also that the CE$_{rich}$
cut ($> 40$) has no relation to N$_{gals} > 65$, so the comparison
between CE and AK-VT has no correspondence to the comparison
between the HMF-VT and AK-VT catalogs.
 
\section{Summary}

This paper presents an intermediate redshift galaxy cluster catalog
($0.1 \lesssim z_{est} \lesssim 0.5$) covering 2,708 $\square^{\circ}$
of DPOSS data, providing a supplement to those presented in papers II and
III.  The catalog is the result of two different density estimators
(AK \& VT) and its combined version contains $\sim$ 10,000 cluster
candidates. To facilitate the selection of subsamples for follow-up
studies we provide both redshift and richness estimates. The redshift
estimates are as accurate as the matched filter redshift output.  By
keeping the contamination rate at 5\%, the probability of
selecting false clusters is minimized. Our
catalog is cleaned post-detection, to discard obvious false
candidates due to bright stars, nearby galaxies and groups, etc.

The main goal of this project is to provide a catalog of rich clusters
to $z_{est} \sim 0.5$ covering a large area of the Northern sky. It is
a valuable source for follow-up studies, 
which can be used to select candidates for lensing studies,
to study the cluster population at intermediate redshifts,
to search for optical rich clusters with low X-ray luminosity, etc.
We plan to carry out a spectroscopic
survey of a subsample of this catalog to confirm the reality of these
systems, as well as deep imaging follow-up.

We present a detailed comparison of the results obtained by the
different algorithms, showing the dependence of the redshift and
richness estimates on cluster centroid, as well as the richness
variation associated to the redshift estimate. We also find a high
level of overlap in the regime where both catalogs are expected to
have high completeness.

Our catalog is also compared to other recent surveys, showing
excellent agreement with those where all catalogs are expected to
have high completeness rates (rich clusters at $0.2 < z_{est} < 0.3$).
However, the cluster catalogs derived from the
higher quality multi-color data of SDSS are more complete than our
catalog. The HMF-VT detects more poor/nearby systems, while the CE
catalog is more complete for higher-$z$ systems. Nonetheless, this
work provides the largest resource for galaxy clusters spanning this
redshift range to date.  It will certainly be superseded in the near
future by the whole sky catalogs generated from SDSS.  However, it
represents the first effort to search for clusters out to the faintest
limits of a sky survey over such a large area.  This cluster catalog
also constitutes an excellent resource for comparison with the final
SDSS catalog, as well as to deep X-ray cluster catalogs.

\acknowledgments

The processing of DPOSS and the production of the Palomar-Norris Sky
Catalog (PNSC) on which this work was based was supported by generous
grants from the Norris Foundation, and other private donors.  Some of
the software development was supported by the NASA AISRP program.  We
also thank the staff of Palomar Observatory for their expert
assistance in the course of many observing runs. We would like to
thank the anonymous referee for useful comments. Finally, we
acknowledge the efforts of the POSS-II team, and the plate scanning
team at STScI. PAAL was supported by the Conselho Nacional de
Desenvolvimento Cient\'{\i}fico e Tecnol\'ogico (CNPq), under
processes 145973/99-9 \& 200453/00-9 and the Funda\c c\~ao de Amparo
\`a Pesquisa do Estado de S\~ao Paulo (FAPESP, process 03/04110-3).
Several undergraduates participated in the data acquisition and
processing towards the photometric calibration of DPOSS. PAAL would
like to thank Massimo Ramella and Walter Boschin for the use of the VT
code, as well as for helpful discussions on the use of this code.  This
project made use of computers from the Center for Advance Computing
Research (CACR), located at Caltech. We are thankful to Roy Williams
and Mark Bartelt for help with using the CACR facilities.  This
research has made use of the NASA/IPAC Extragalactic Database (NED)
which is operated by the Jet Propulsion Laboratory, California
Institute of Technology, under contract with the National Aeronautics
and Space Administration.

\appendix

\clearpage


\begin{figure}
\plotone{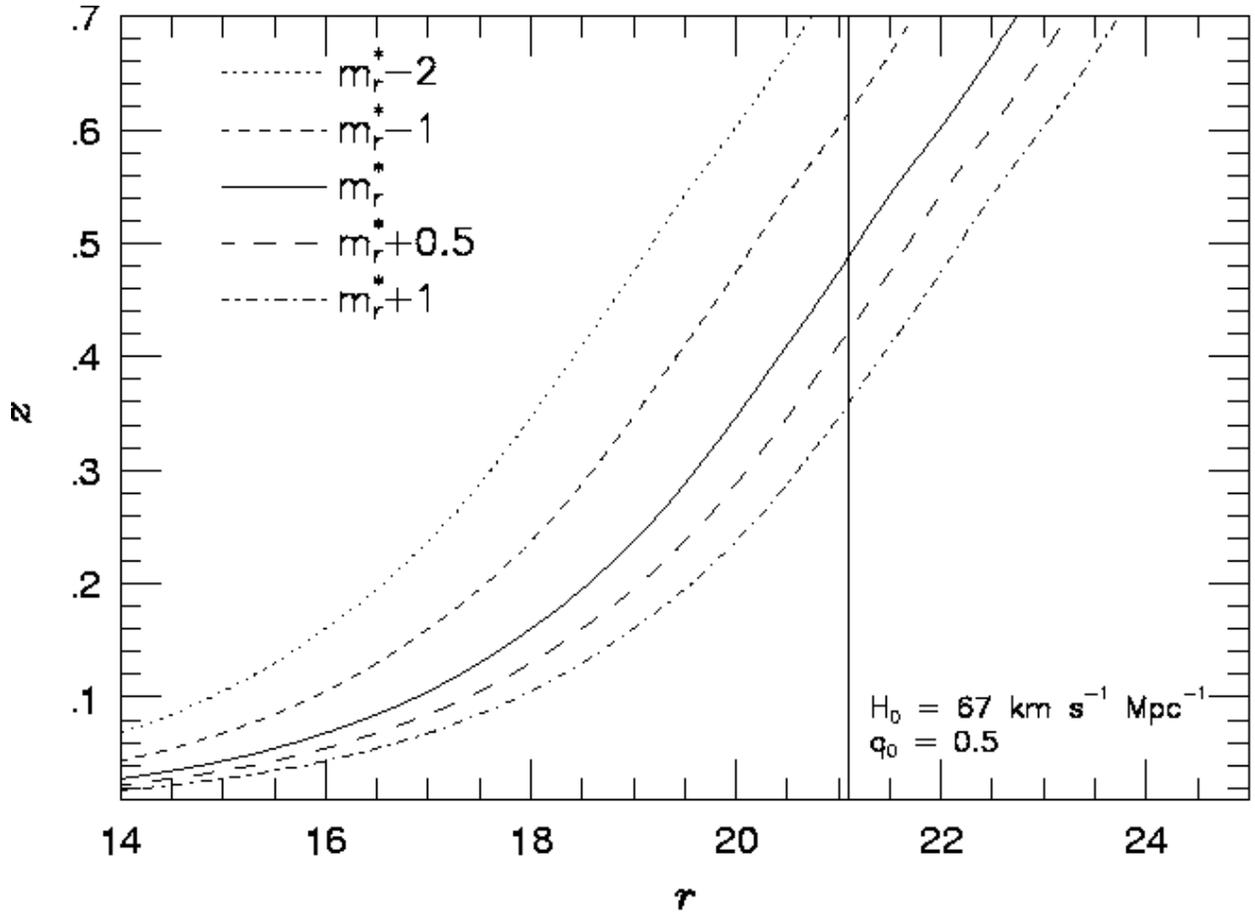}
\caption{{The magnitude-redshift relation for elliptical galaxies
with $M^*_r = -21.52$. The survey limit is indicated by the vertical line
at $m_r$ = 21.1. \label{fig1}}}
\end{figure}

\clearpage

\begin{figure}
\plotone{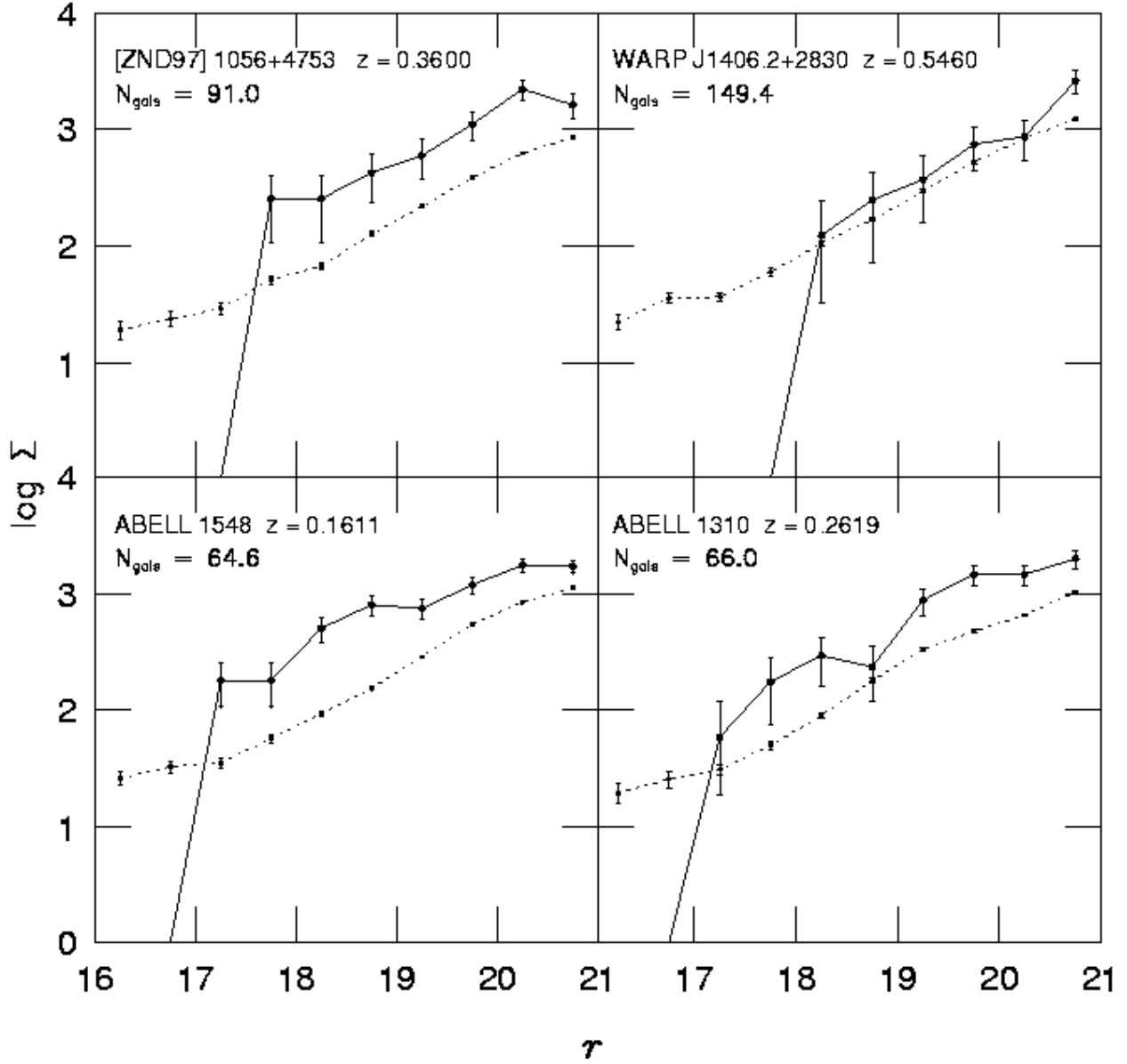}
\caption{{Differential magnitude distributions for four rich
galaxy clusters with known spectroscopic redshifts.
Cluster counts are shown as solid
lines, while dotted lines show the local background counts. All counts
are normalized to an area of one square degree. \label{fig2}}}
\end{figure}

\clearpage

\begin{figure}
\plotone{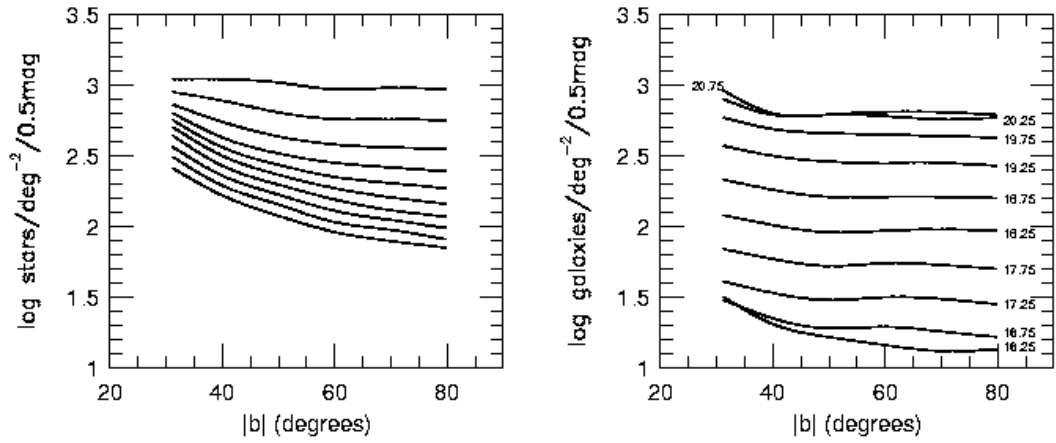}
\caption{{Star and galaxy counts as a function of galactic latitude
for different magnitude bins. The left panel shows the star counts, while
the right panel shows the galaxy counts. The magnitude bins are indicated
in the right panel. \label{fig3}}}
\end{figure}

\clearpage

\begin{figure}
\plotone{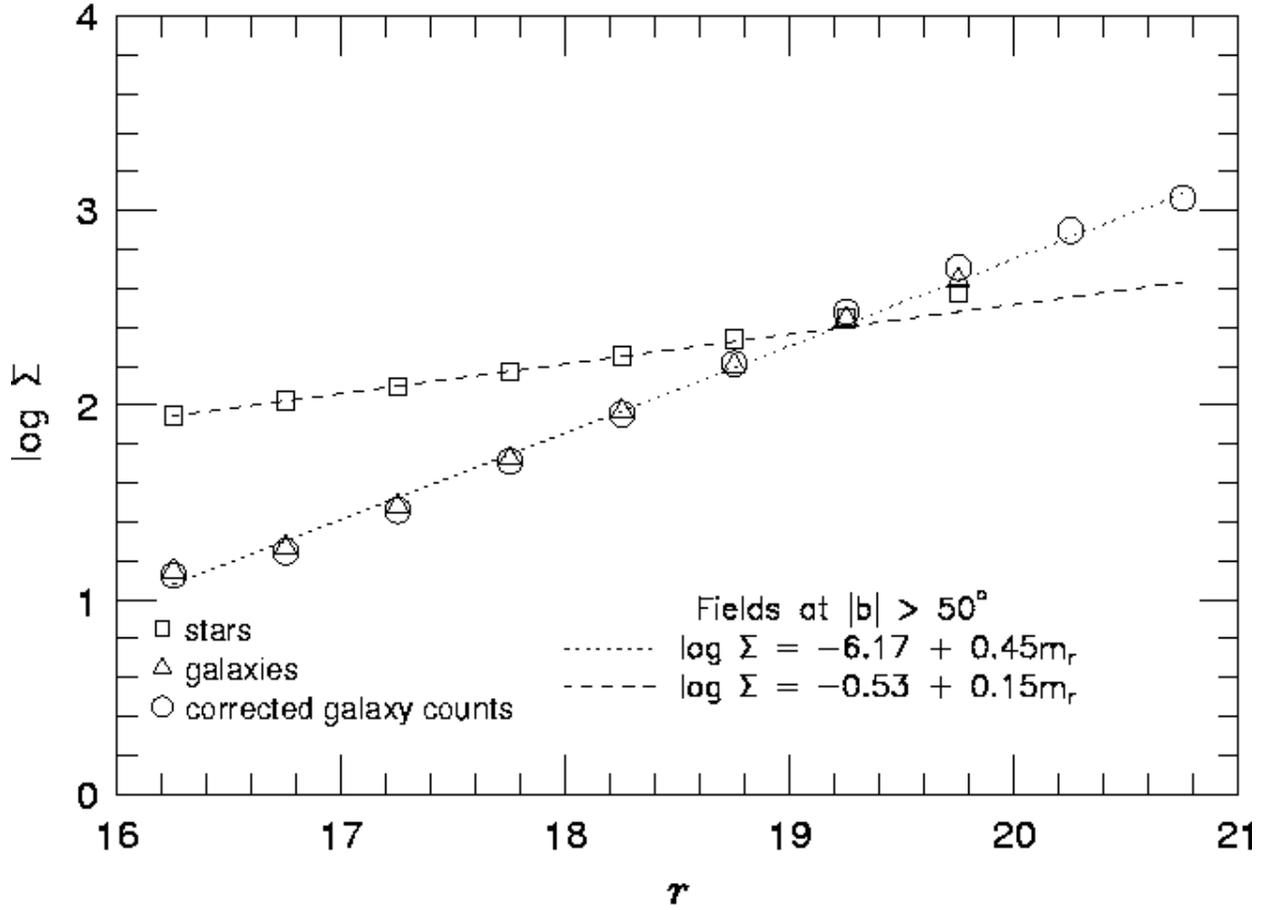}
\caption{{Star and galaxy counts
at $|b| > 50^\circ$. Stars are shown as squares and galaxies
as triangles. The dotted line is the best fit to the bright galaxy counts
extrapolated to $m_r = 21$. The dashed line represents the bright
star counts ($m_r \le 18.5$) extrapolated to $m_r = 21$.
Circles represent the corrected galaxy counts out to
$m_r = 21$. Error bars (not shown here for clearness) are at the same level of 
the error estimates
for field counts in Figure 2. \label{fig4}}}
\end{figure}

\clearpage

\begin{figure}
\plotone{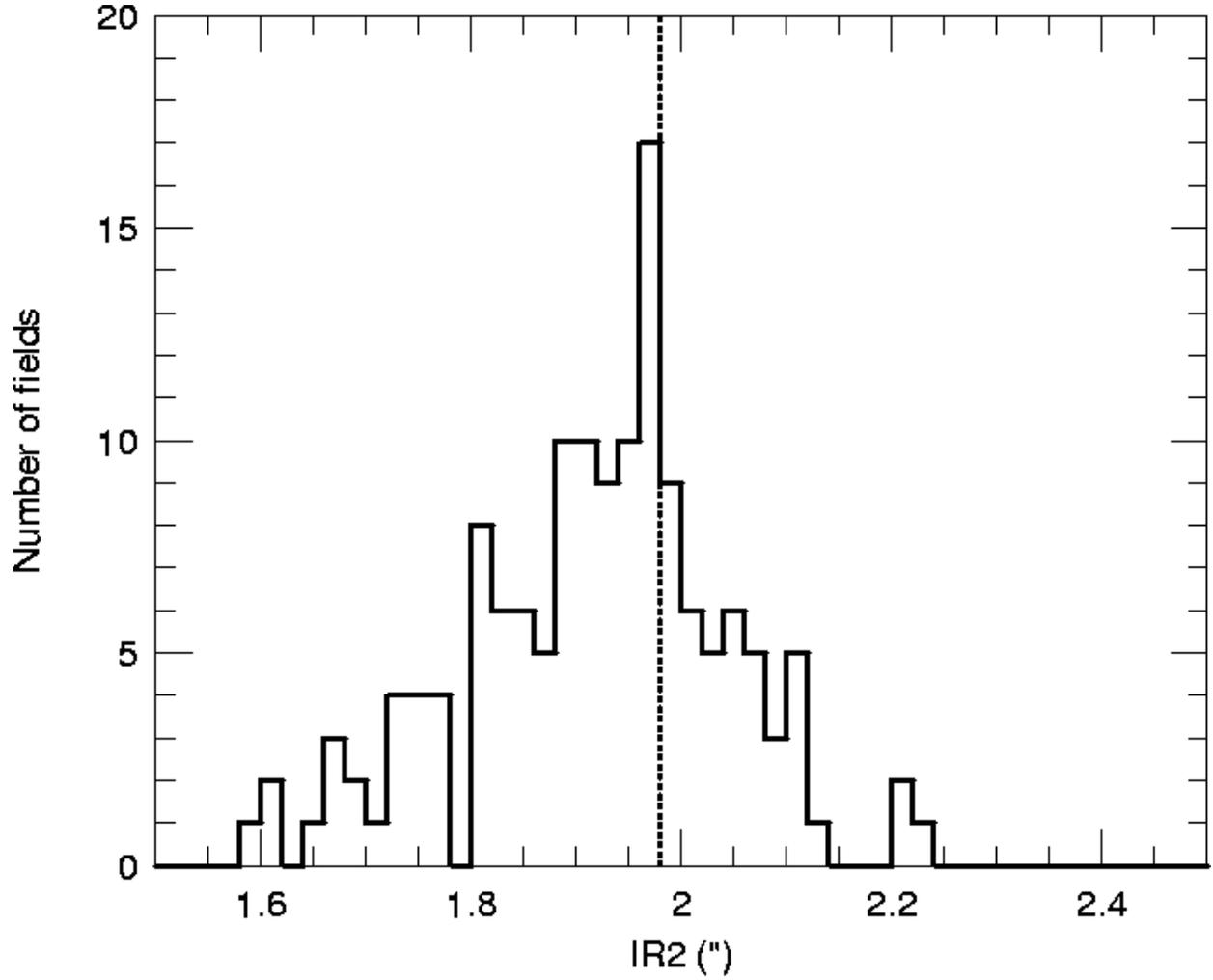}
\caption{{The intensity-weighted second moment of the light distribution for
stellar objects (IR2) in the $r$-band is given as an indication of the
quality of DPOSS fields. Its distribution is shown above for
the well calibrated and high latitude DPOSS fields. We select 
fields with IR2 $ < 1.98''$. \label{fig5}}}
\end{figure}

\clearpage

\begin{figure}
\epsscale{0.70}
\plotone{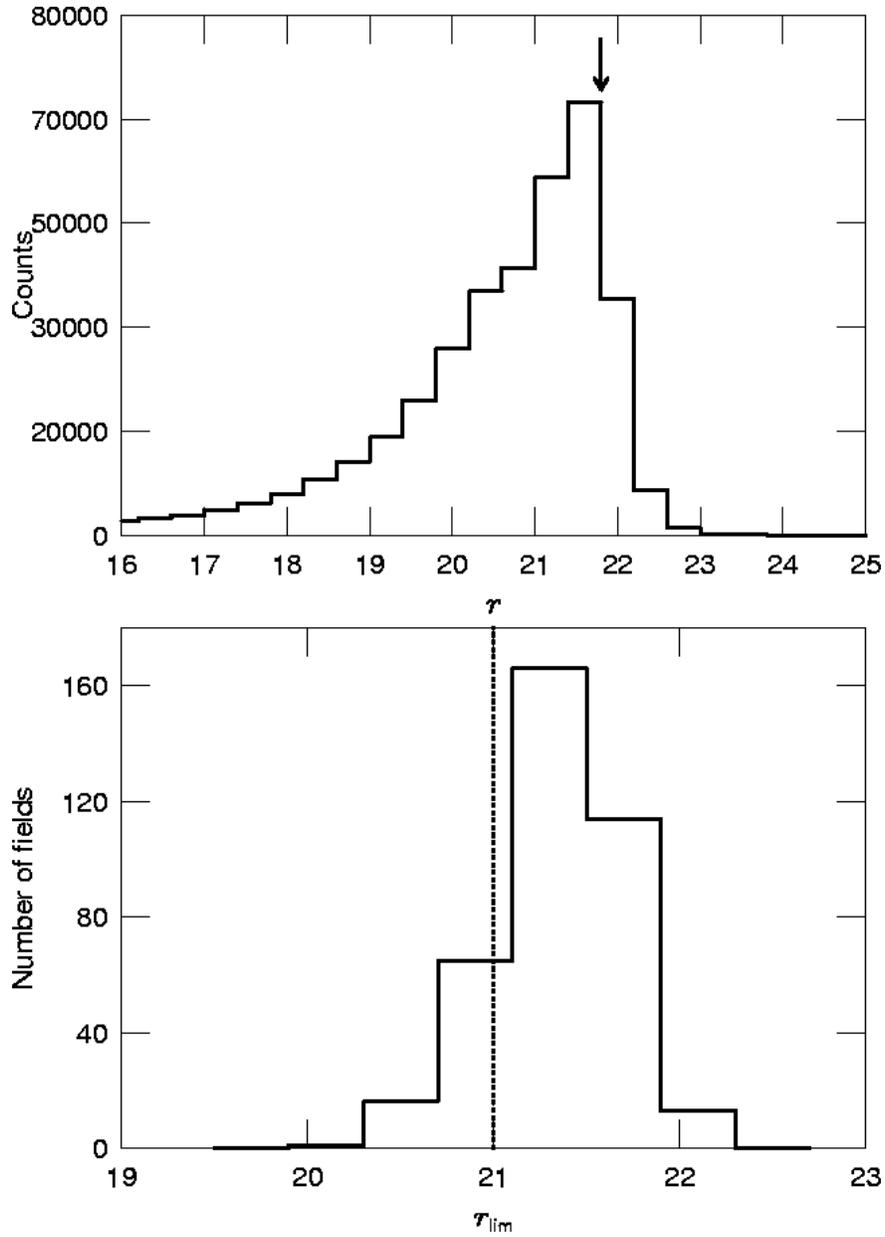}
\caption{{Typical magnitude distribution of a DPOSS field (top). The
limiting magnitude is indicated by the arrow. In the bottom panel
we show the distribution of limiting magnitudes for 375 DPOSS plates.
We exclude fields with $m_{r,lim} < 21$.
\label{fig6}}}
\epsscale{1.0}
\end{figure}

\clearpage

\begin{figure}
\plotone{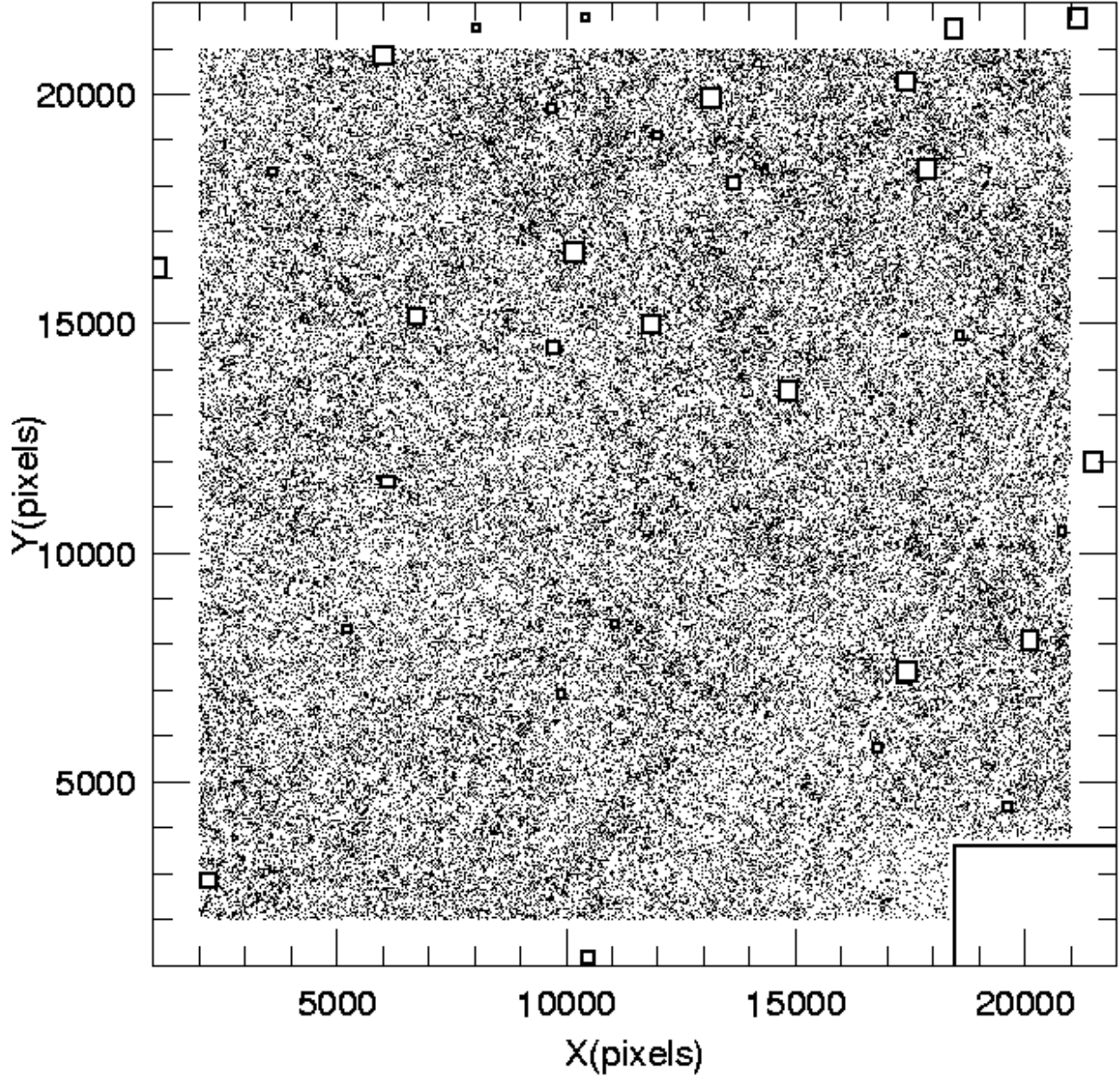}
\caption{{Galaxy distribution in DPOSS Field 444 ($\alpha = 201.64^{\circ}, ~\delta = 29.74^{\circ}$)
for $16.0 \le m_r \le 21.1$. Excised areas due to bright
objects are indicated by rectangles. A cleaned path due to a satellite trail
is visible in the lower right corner.
\label{fig7}}}
\end{figure}

\clearpage

\begin{figure}
\plotone{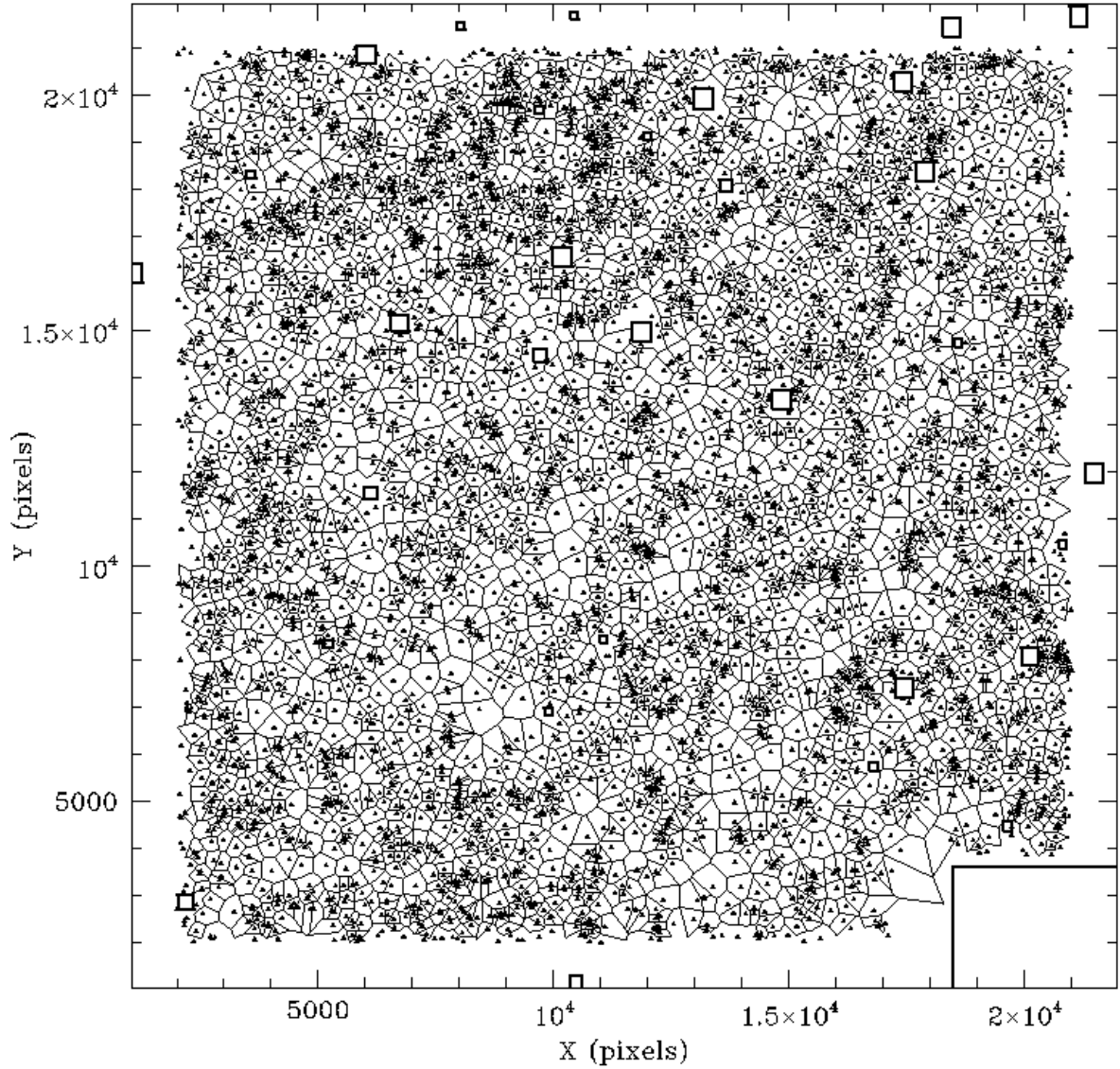}
\caption{{Voronoi Tessellation of galaxies with $17.0 \le m_r \le 18.5$ in 
Field 444. Each triangle represents a
galaxy surrounded by its associated Voronoi cell (indicated by
the polyhedrals).
Excised areas (due to bright objects) are shown as rectangles.
\label{fig8}}}
\end{figure}

\clearpage

\begin{figure}
\plotone{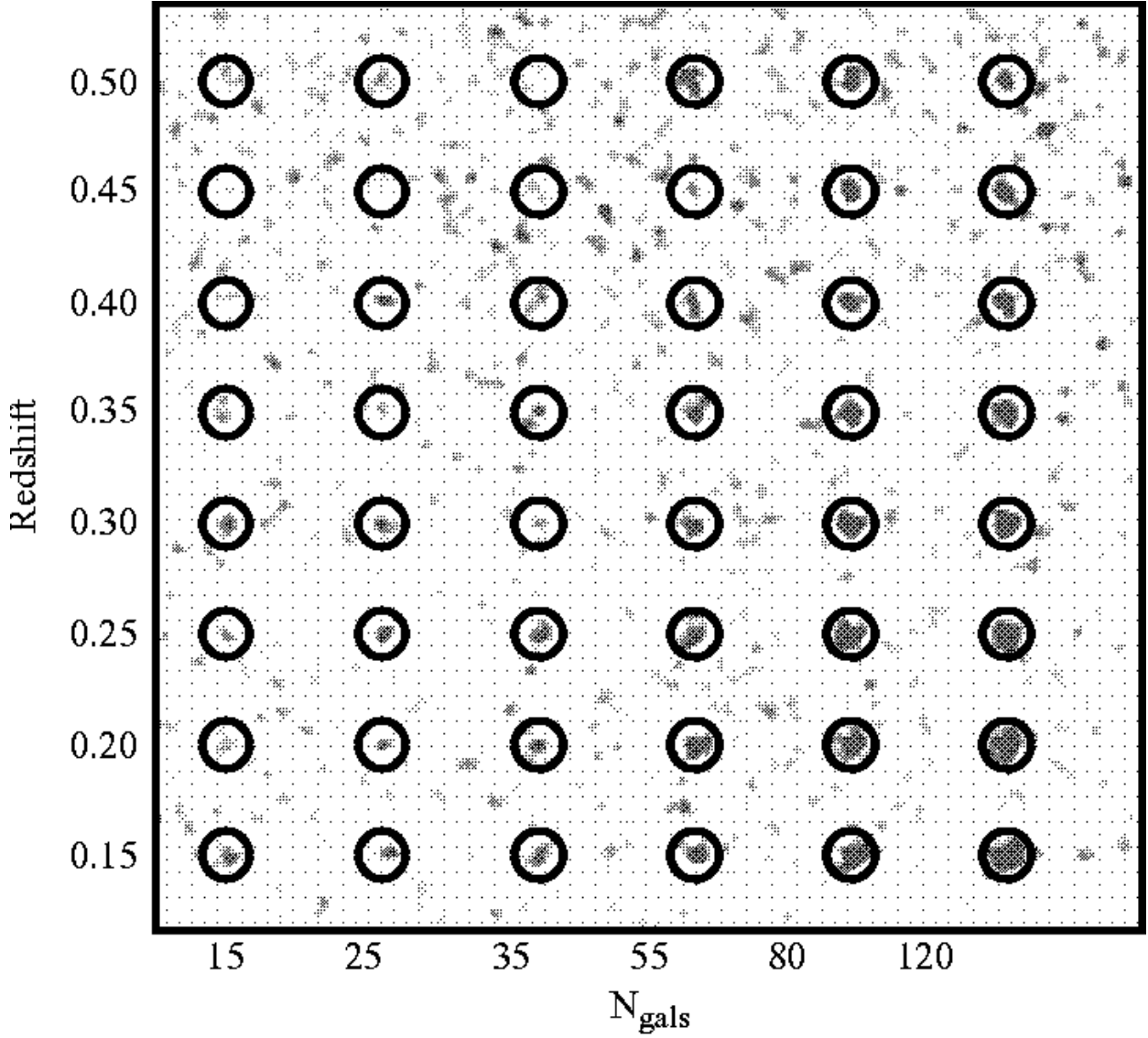}
\caption{{Density map of a simulated background galaxy
distribution with 48 artificial clusters inserted.
The initial kernel adopted has a 260$''$ radius. Artificial clusters
are marked by circles. Each row represents a different
redshift ($z = $0.15, 0.20, 0.25, 0.30, 0.35, 0.40, 0.45, 0.50),
increasing from bottom to top. Richness increases from left
to right (N$_{gals} = 15, 25, 35, 55, 80, 120$). \label{fig9}}}
\end{figure}

\clearpage

\begin{figure}
\epsscale{0.70}
\plotone{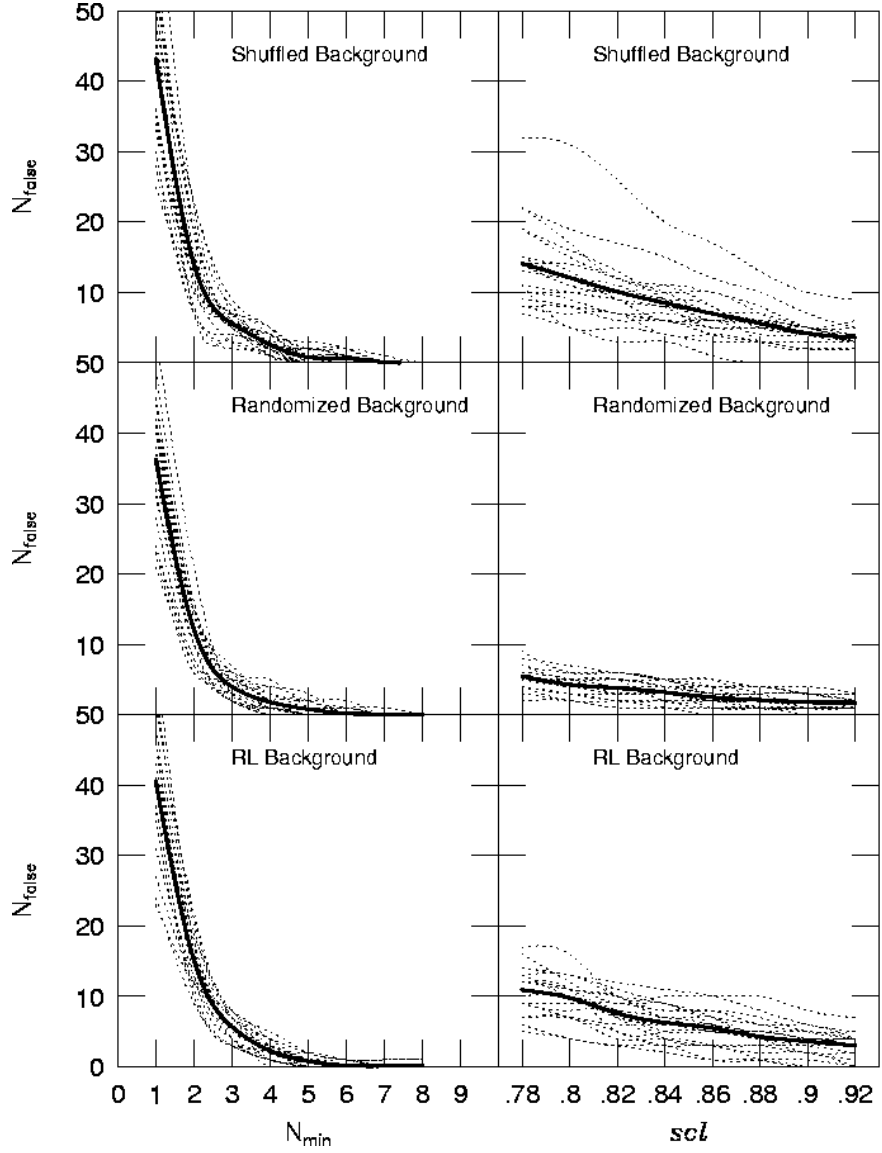}
\caption{{Comparison of the number of false detections using an RL
background distribution, a randomized background, and a shuffled
background. The dotted lines show the variation of N$_{false}$ with
N$_{min}$ (left panels) and {\it\bf scl} (right panels) for 20 
different DPOSS fields, while the solid line shows the mean
variation. \label{fig10}}}
\epsscale{1.0}
\end{figure}

\clearpage

\begin{figure}
\epsscale{0.70}
\plotone{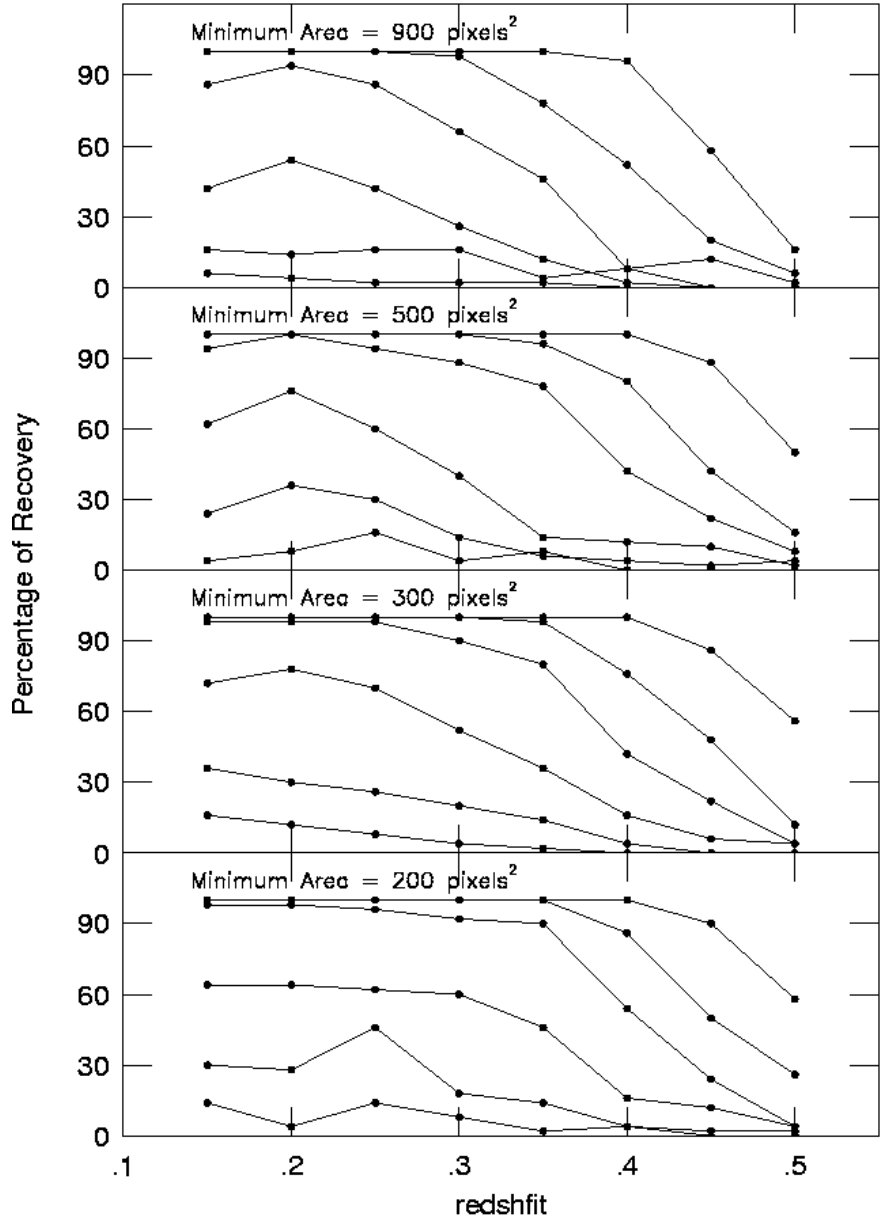}
\caption{{ The selection function with the AK code for
4 different values of the minimum area: 200, 300, 500
and 900 pixels$^2$. Within each panel, the lines
represent different richness (N$_{gals} = 15,
25, 35, 55, 80, 120$), increasing from bottom to top. \label{fig11}}}
\epsscale{1.0}
\end{figure}

\begin{figure}
\plotone{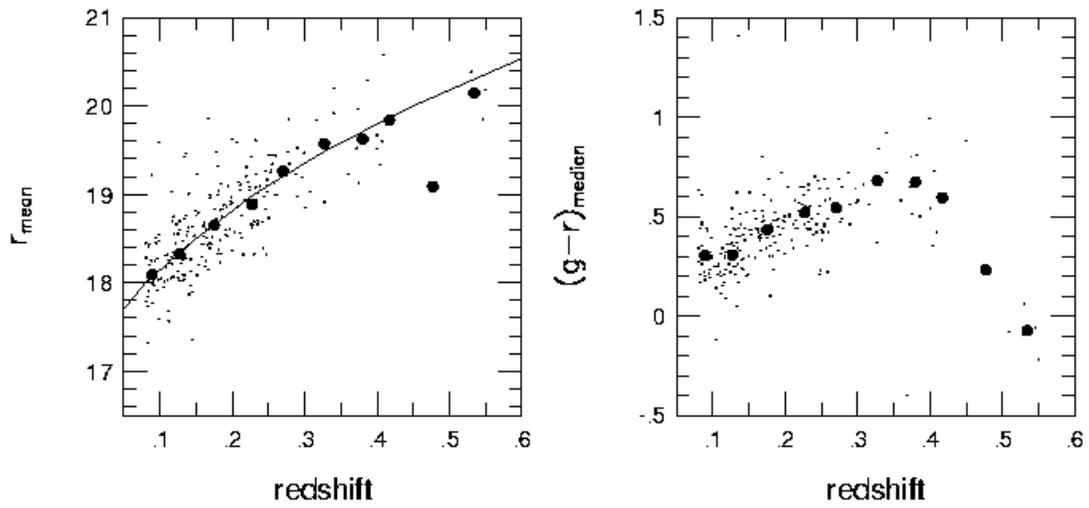}
\caption{{ The mean galaxy magnitude $r_{mean}$ and median color
$(g-r)_{median}$ for clusters as a function of spectroscopic
redshift. The solid line shown in the left panel represents the
empirical relation used to estimate redshifts for this
work. \label{fig12}}}
\end{figure}

\clearpage

\begin{figure}
\plotone{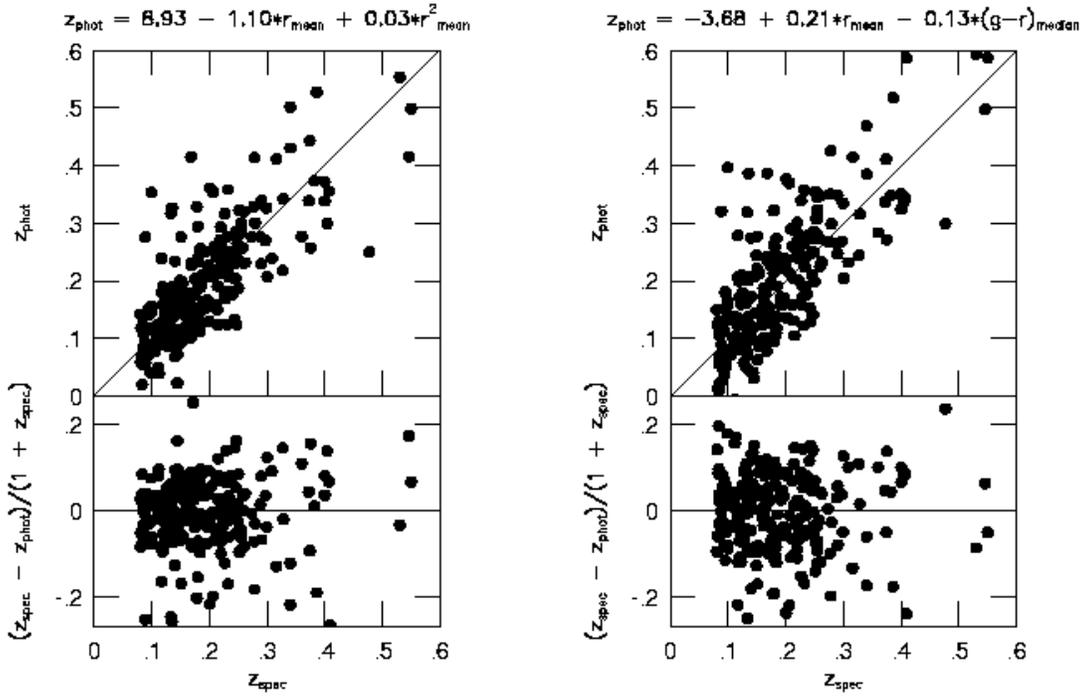}
\caption{{The photometric redshift estimate {\it vs.} spectroscopically
measured redshift for 238 clusters. Residuals as a function of magnitude
are shown in the bottom panel of each plot. We show
photometric estimates based on only the mean magnitude (left panel), as well
as both the mean magnitude and median color $(g-r)$ (right panel).
\label{fig13}}}
\end{figure}

\clearpage

\begin{figure}
\epsscale{0.70}
\plotone{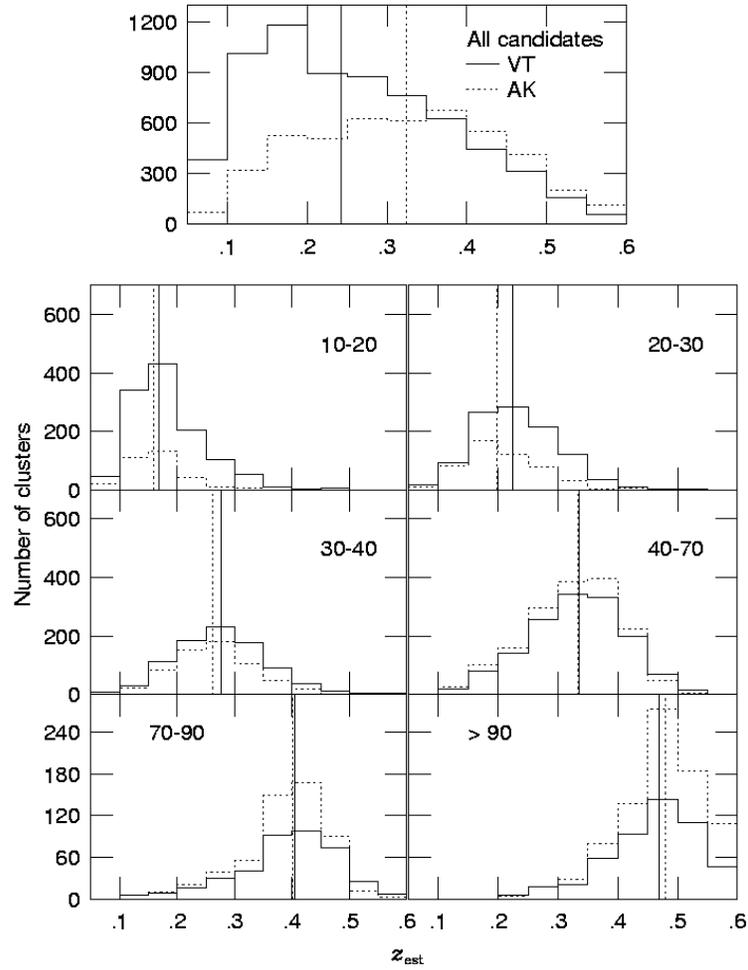}
\caption{{The estimated redshift distribution in different richness ranges and for
the whole sample (top panel). The richness (N$_{gals}$) range
is indicated in all panels. The median redshift is also
indicated for the VT (solid line) and AK (dotted line) candidates.
\label{fig14}}}
\epsscale{1.0}
\end{figure}

\clearpage

\begin{figure}
\epsscale{0.70}
\plotone{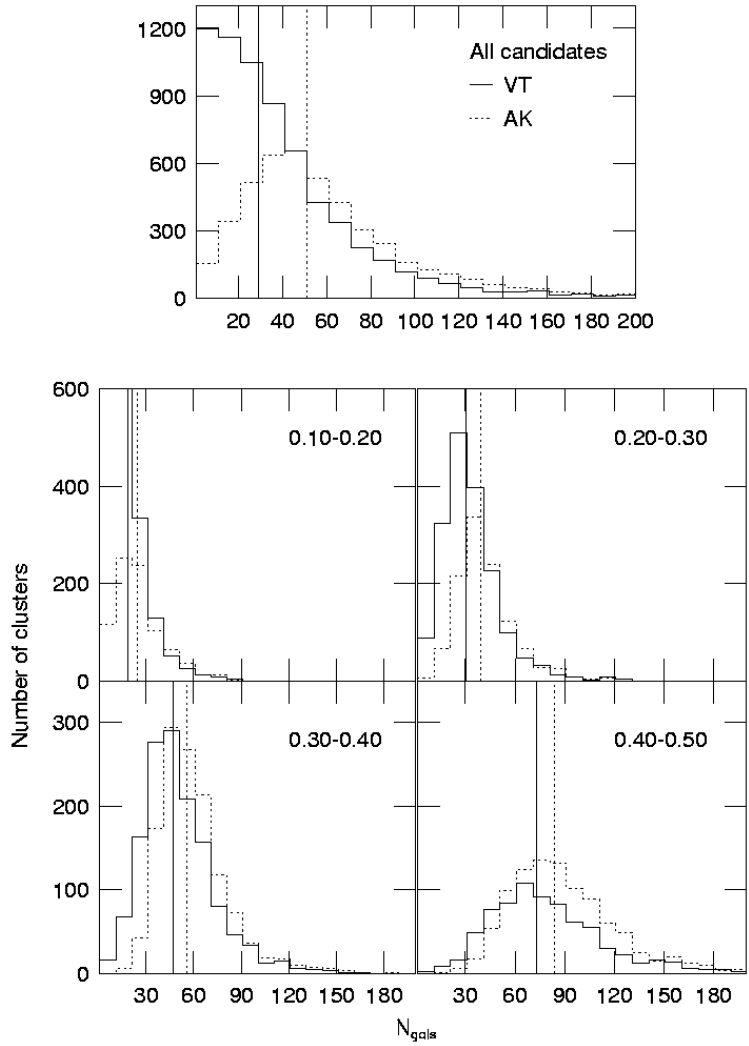}
\caption{{The richness distribution in bins of estimated redshift
and for the whole sample (top panel). The redshift bins
are indicated in each panel, as well as the median richness,
which is indicated by the vertical lines (solid for the VT candidates and
dotted for the AK). \label{fig15}}}
\epsscale{1.0}
\end{figure}

\clearpage

\begin{figure}
\plotone{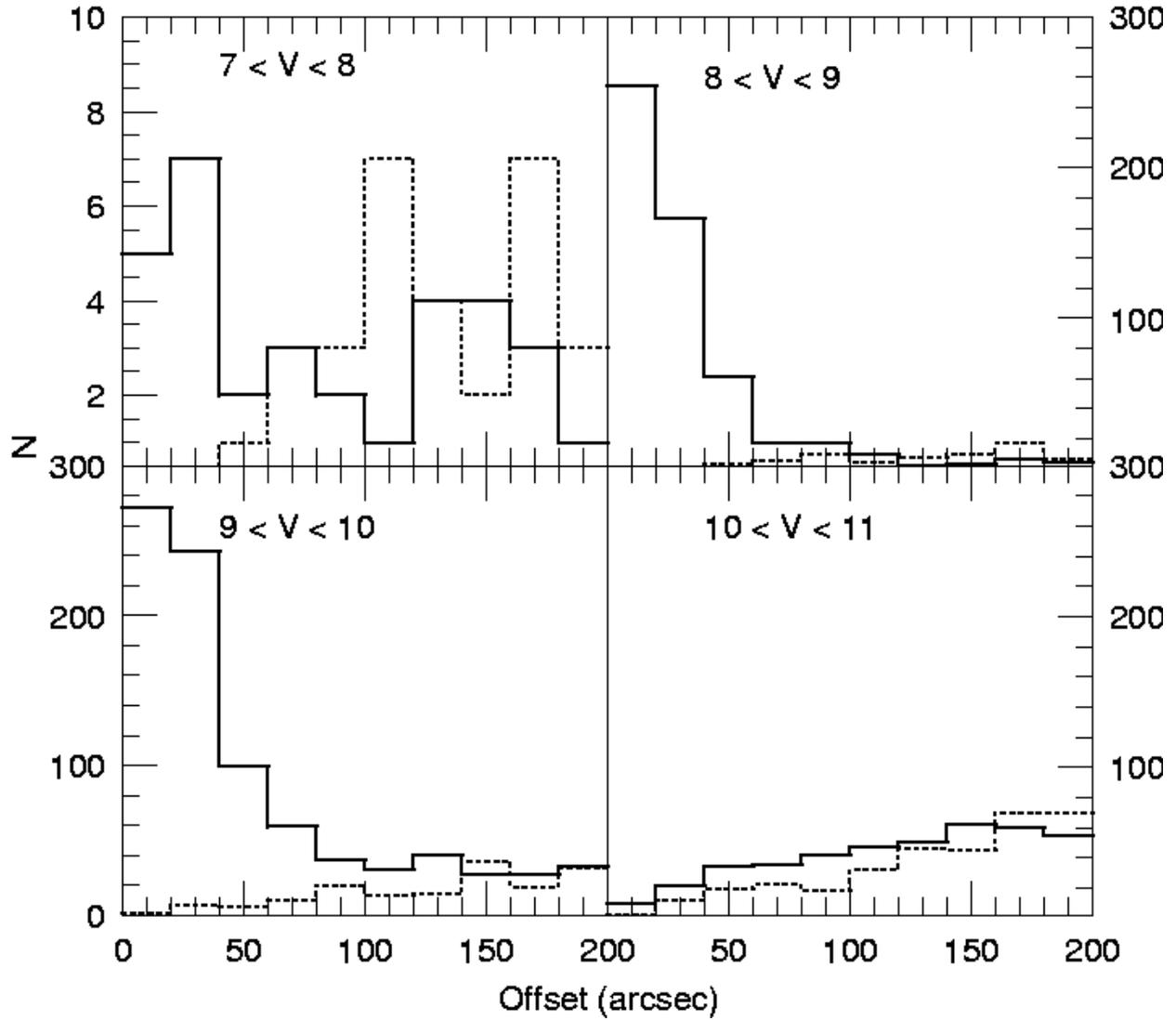}
\caption{{Offset distribution between Tycho-2 stars and 
cluster candidates (solid lines). The dotted lines 
represent the offset distribution using a mock star catalogue. 
The magnitude ranges are indicated on each panel. \label{fig16}}}
\end{figure}

\clearpage

\begin{figure}
\plotone{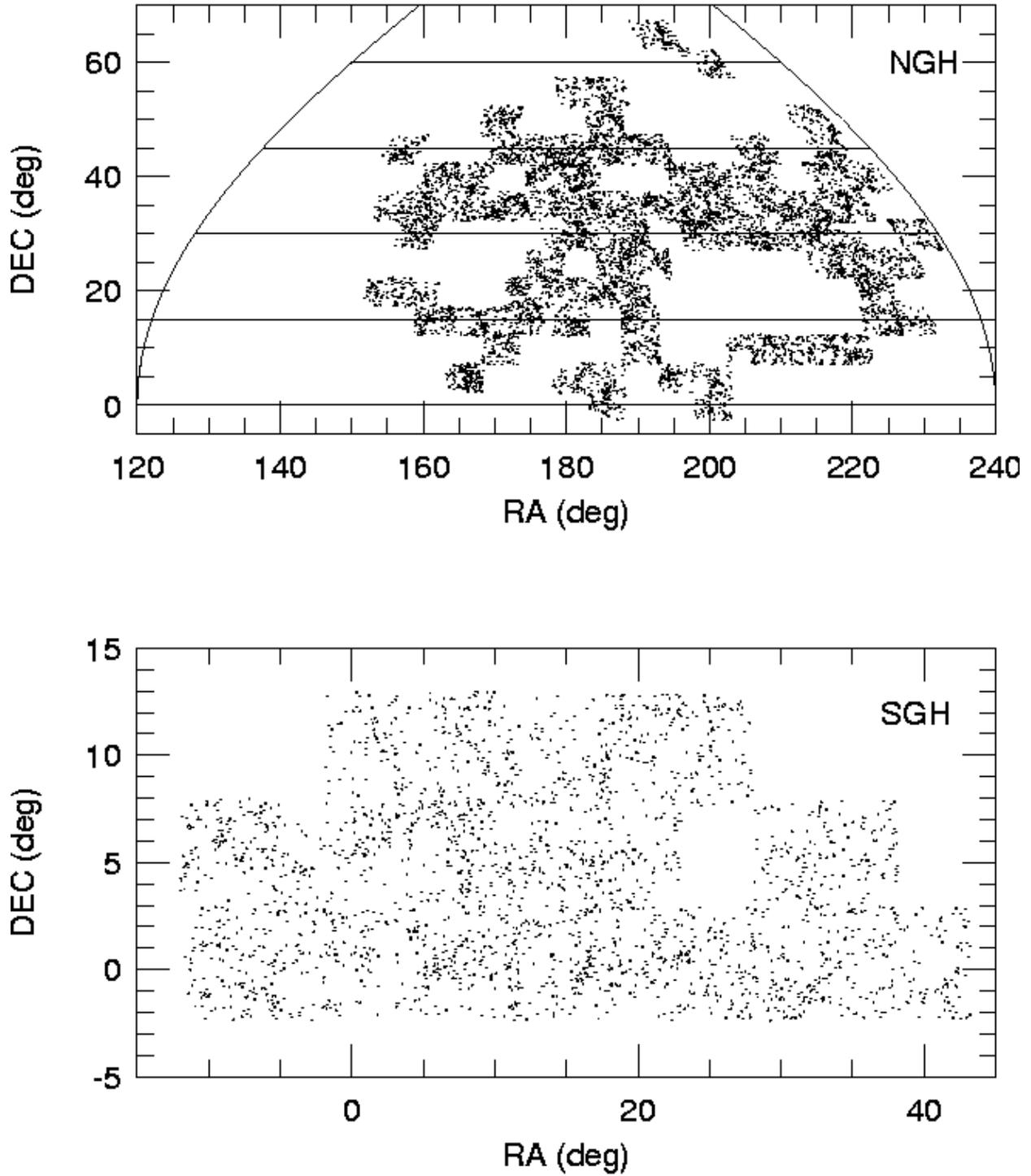}
\caption{{The sky distribution in equatorial coordinates for the
combined AK-VT catalog (9,956 candidates). The region covering
the northern galactic hemisphere is shown in the top panel, while
the bottom panel shows the southern galactic hemisphere. 
\label{fig17}}}
\end{figure}

\clearpage

\begin{figure}
\plotone{lopes.fig18.epsi}
\caption{{Examples of intermediate redshift rich clusters candidates
detected in DPOSS. Images are 250$''$ on a side. NSCS J145302+580319 =
Abell 1995 at $z_{spec} = 0.32$; NSCS J110313+041937 was not
previously catalogued; NSCS J121732+364150 = RX J1217.5+3641; NSCS
J095850+334539 = ZwCl 0955.8+3401. Matches were found in NED using a
search radius of 3 arcmin. \label{fig18}}}
\end{figure}

\clearpage

\begin{figure}
\plotone{lopes.fig19.epsi}
\caption{{Candidate NSCS J234730-000853, also found by the CE
method \citep{got02}, with an estimated redshift from SDSS data
of $z_{est} = 0.24$. 
The left panel shows a DPOSS F-plate image, while the right panel shows
a CCD image taken at the Palomar 60$''$ telescope. Images are
250$''$ on a side. This figure demonstrates the difficulty of visually
confirming distant rich clusters from DPOSS plates.\label{fig19}}}
\end{figure}

\clearpage

\begin{figure}
\plotone{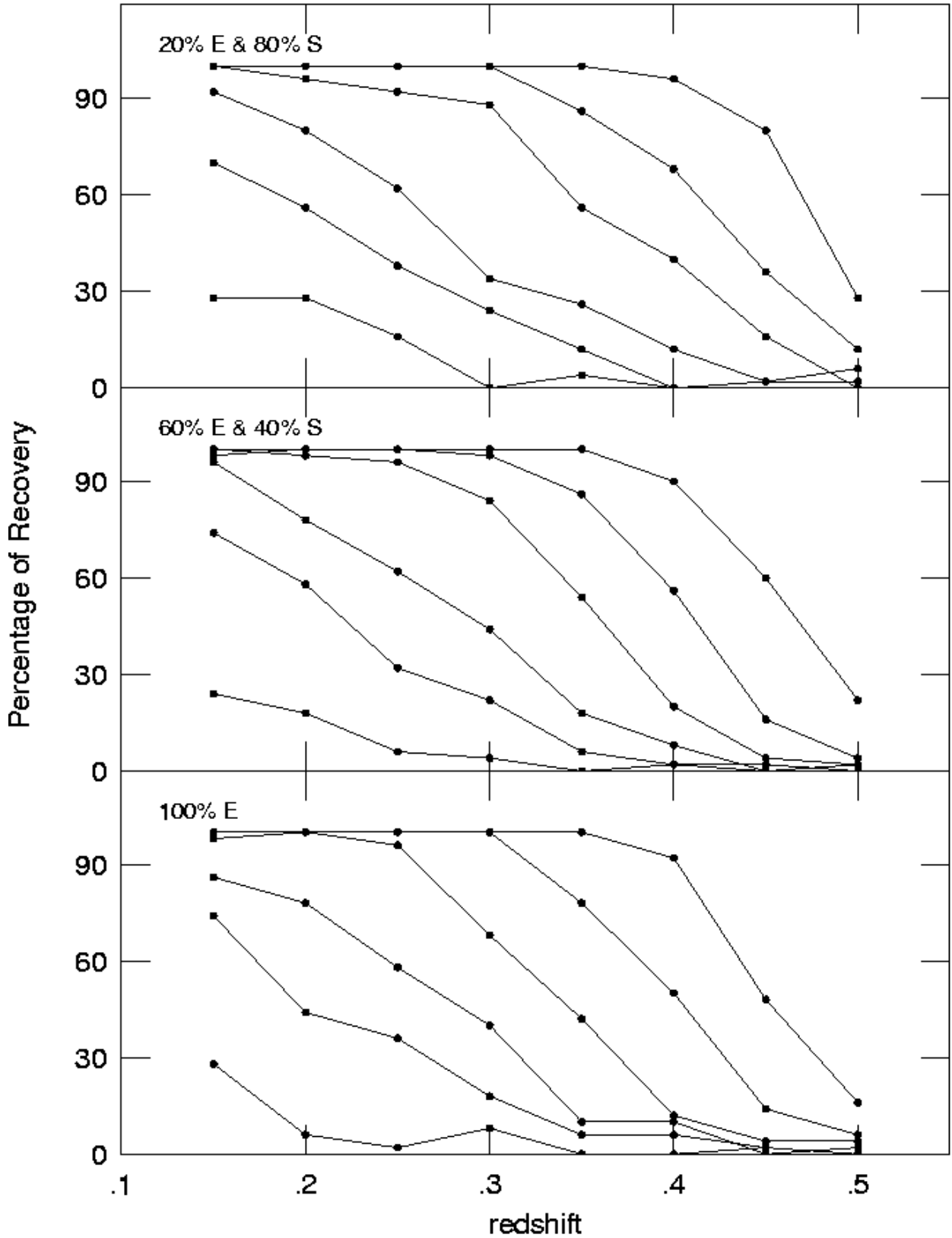}
\caption{{Comparison of the selection function evaluated with three
different cluster compositions. The top panel shows the
results for 20\% E \& 80\% Sbc, the middle panel is for
60\% E \& 40\% Sbc, while the bottom panel is for 100\%
 ellipticals. The richness classes are as indicated in Figure 11.
\label{fig20}}}
\end{figure}

\clearpage

\begin{figure}
\plotone{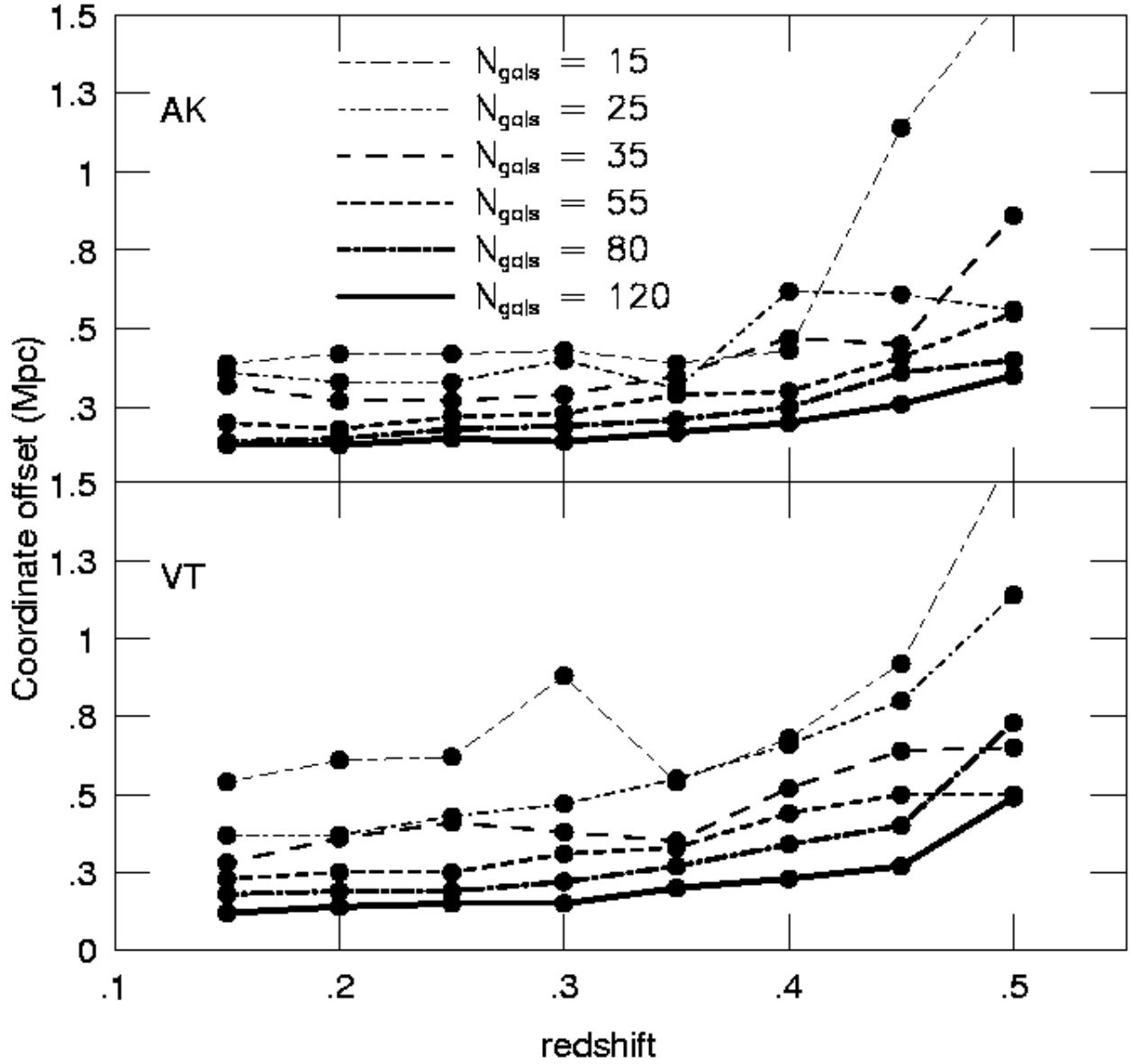}
\caption{{Positional offsets (in Mpc) for the VT and AK codes
as a function of redshift. Each line represent a different richness class
(N$_{gals} = 15, 25, 35, 55, 80, 120$), with the thickest representing
the richest clusters.\label{fig21}}}
\end{figure}

\clearpage

\begin{figure}
\plotone{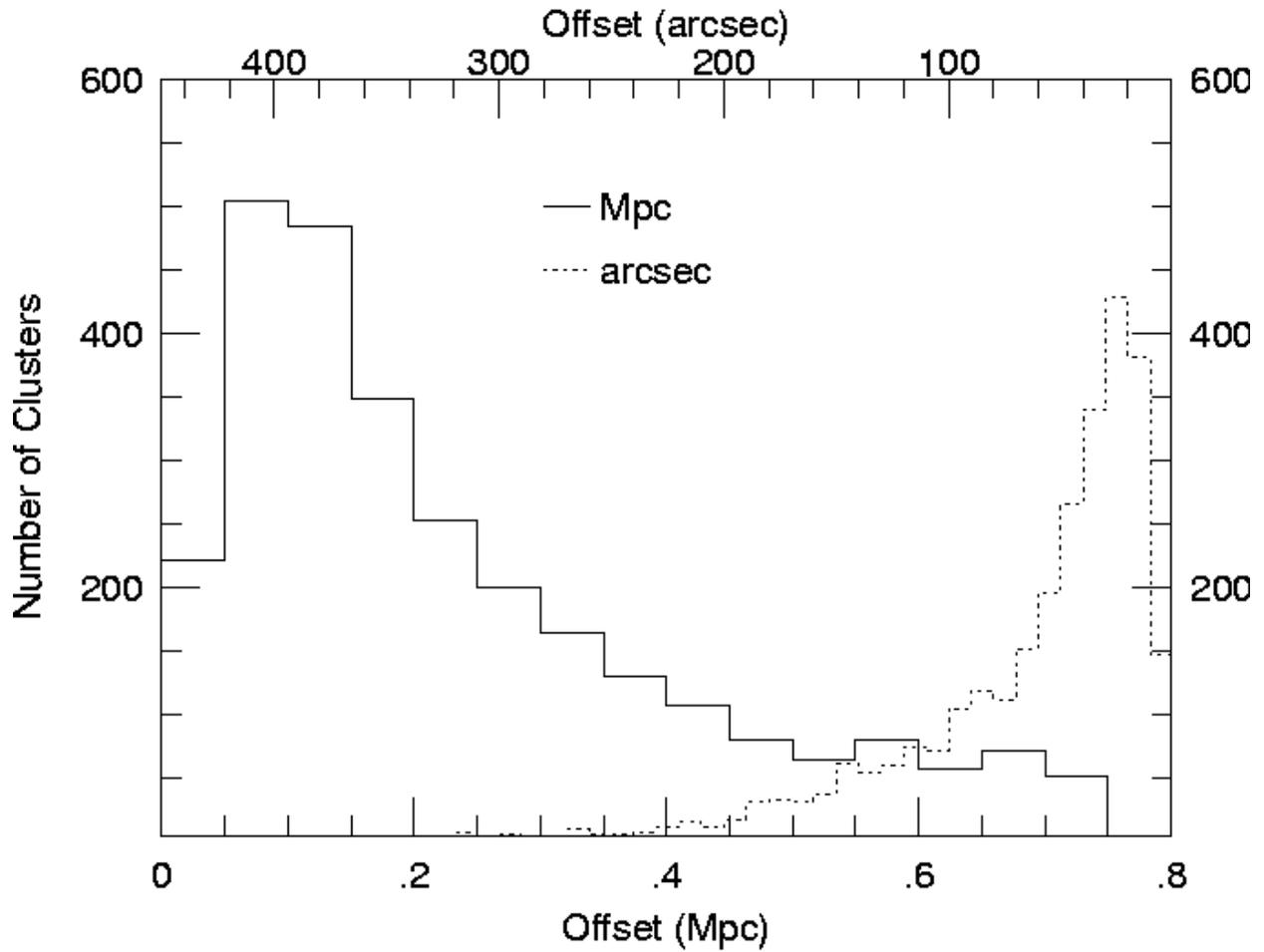}
\caption{{Centroid offset distribution for common clusters from
the VT and AK codes. Offsets in Mpc are indicated by
the solid line, while the offsets in arcsec are
represented by the dotted line. \label{fig22}}}
\end{figure}

\clearpage

\begin{figure}
\plotone{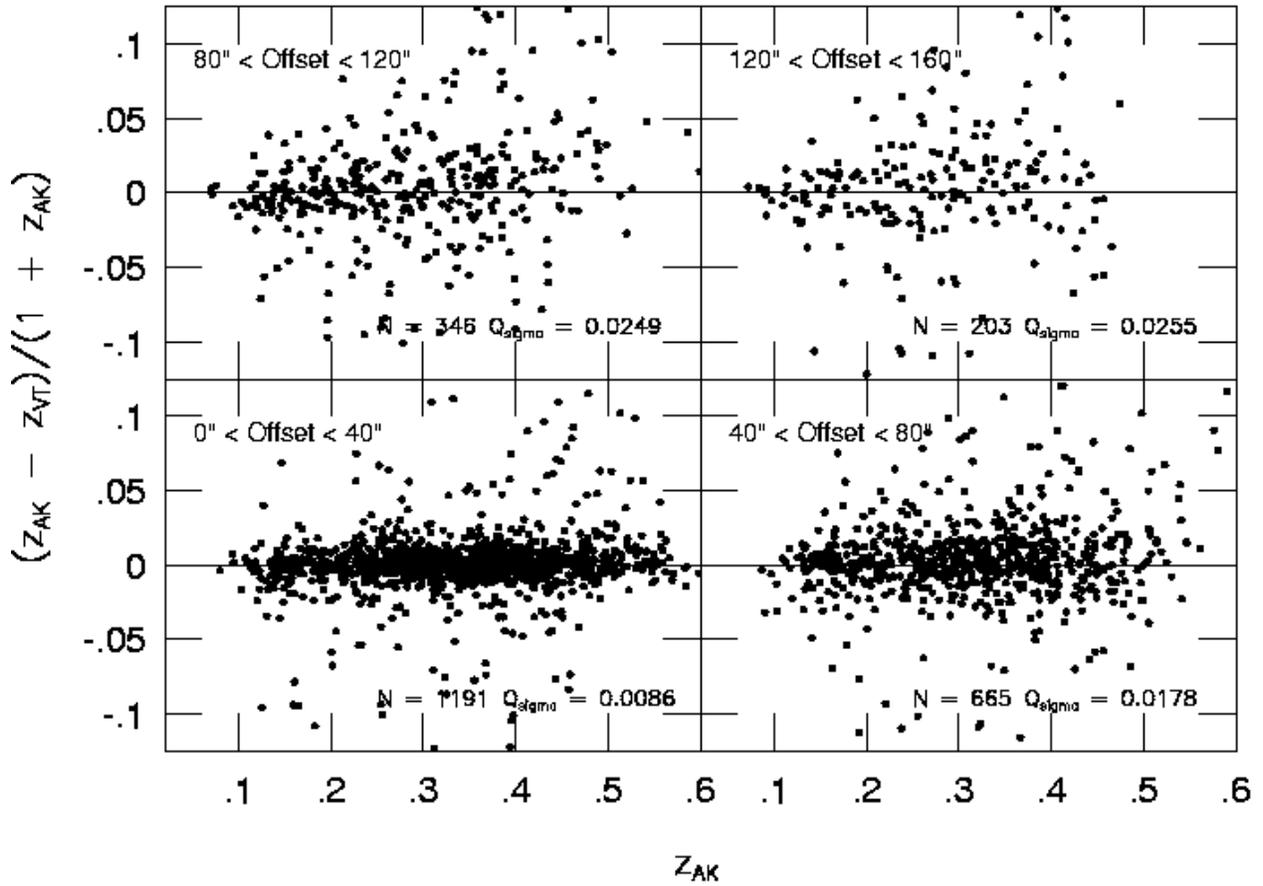}
\caption{{The dependence of the redshift estimate
on the cluster centroid. Redshift residuals as
a function of redshift are shown for different offset
ranges, which are indicated on each panel. The number of
clusters (N) is also indicated.
\label{fig23}}}
\end{figure}

\clearpage

\begin{figure}
\plotone{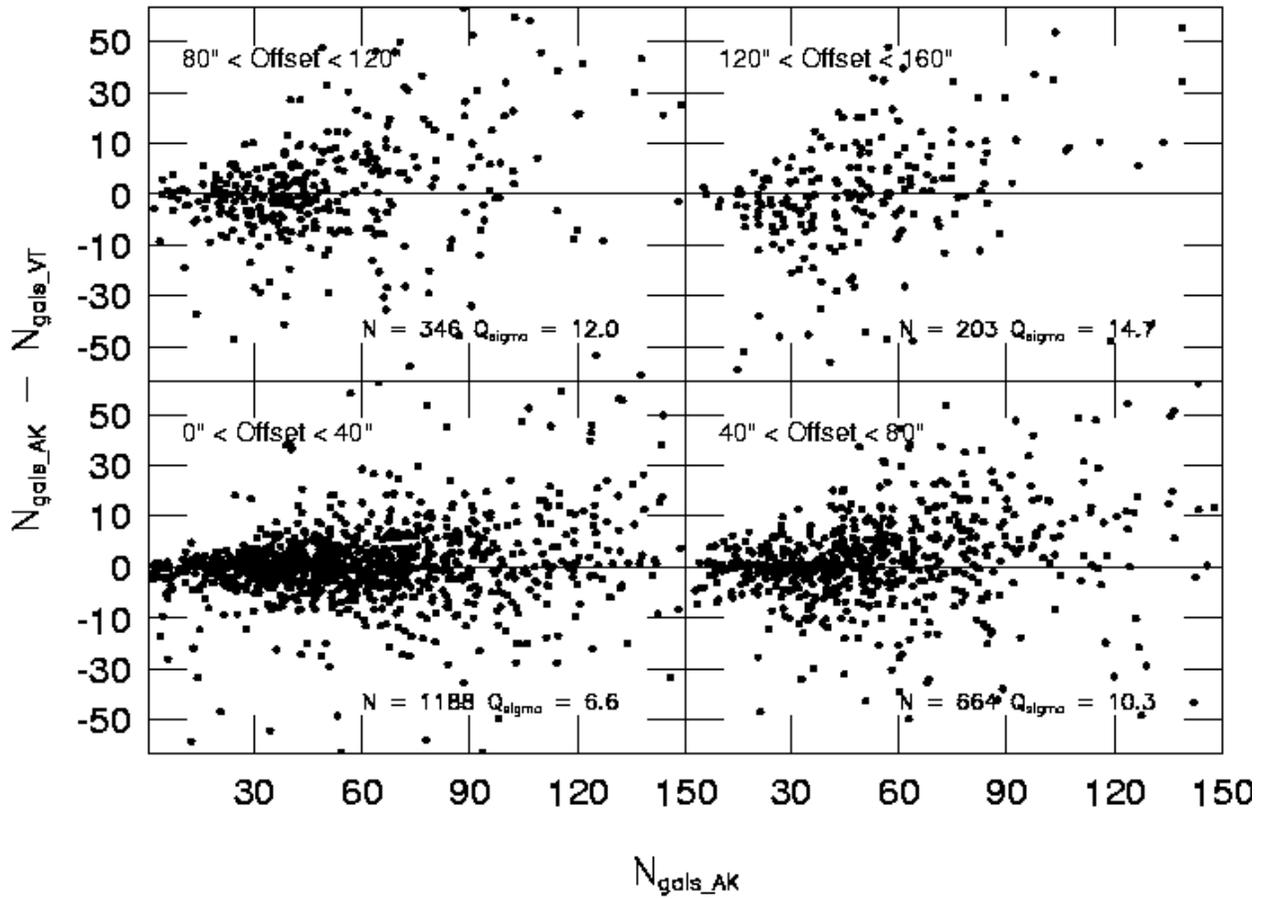}
\caption{{Richness residuals as a function of richness
are shown for different offset
ranges, which are indicated on each panel. The number of
clusters (N) is also indicated. \label{fig24}}}
\end{figure}

\clearpage

\begin{figure}
\plotone{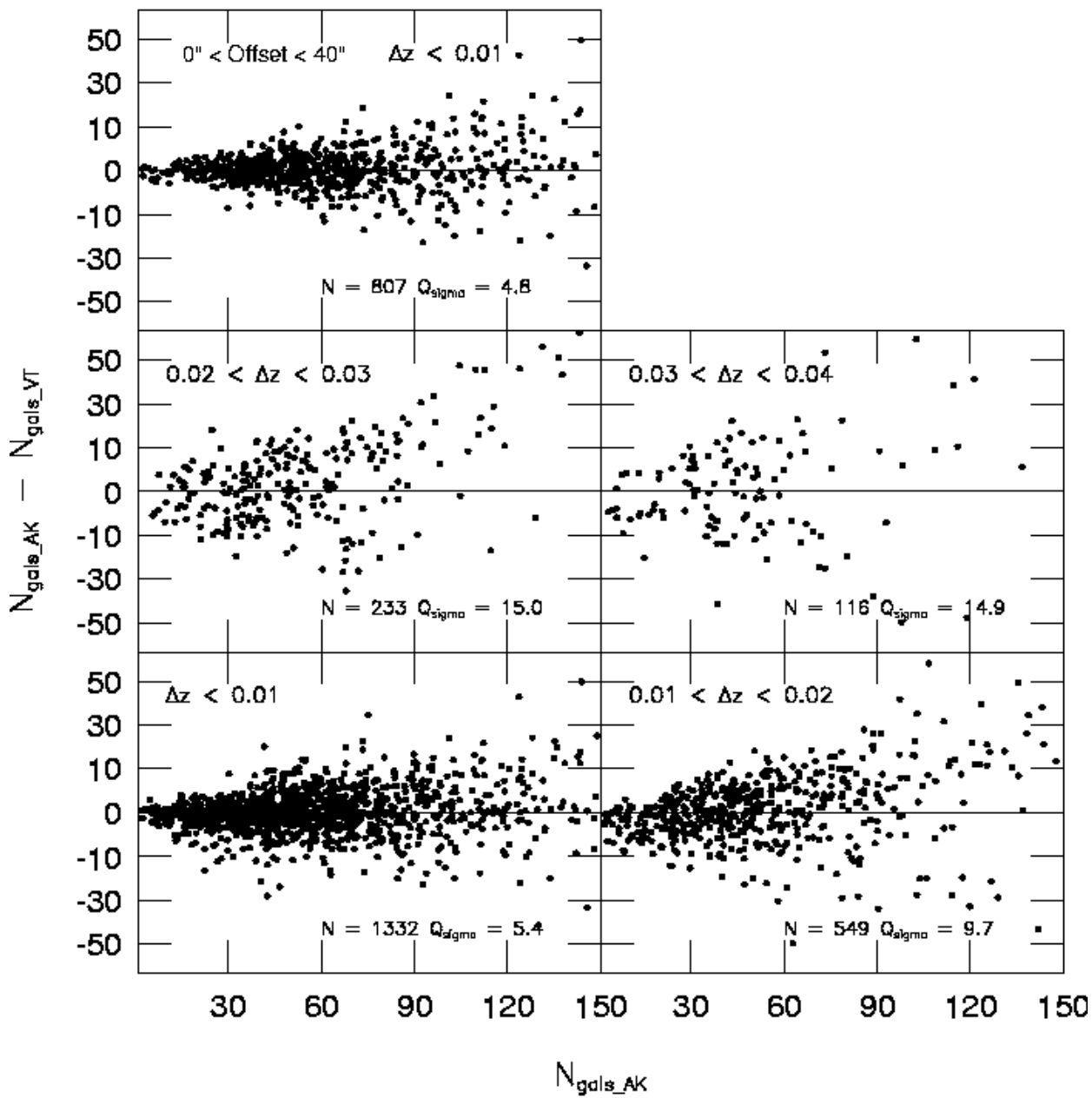}
\caption{{Richness residuals as a function of richness
are shown for different bins of redshift residuals (indicated
on each panel). The number of clusters (N) is also indicated.
The upper left panel shows the results
when selecting only clusters with offsets $< 40''$ and
$\Delta z < 0.01$.
\label{fig25}}}
\end{figure}

\clearpage

\begin{figure}
\plotone{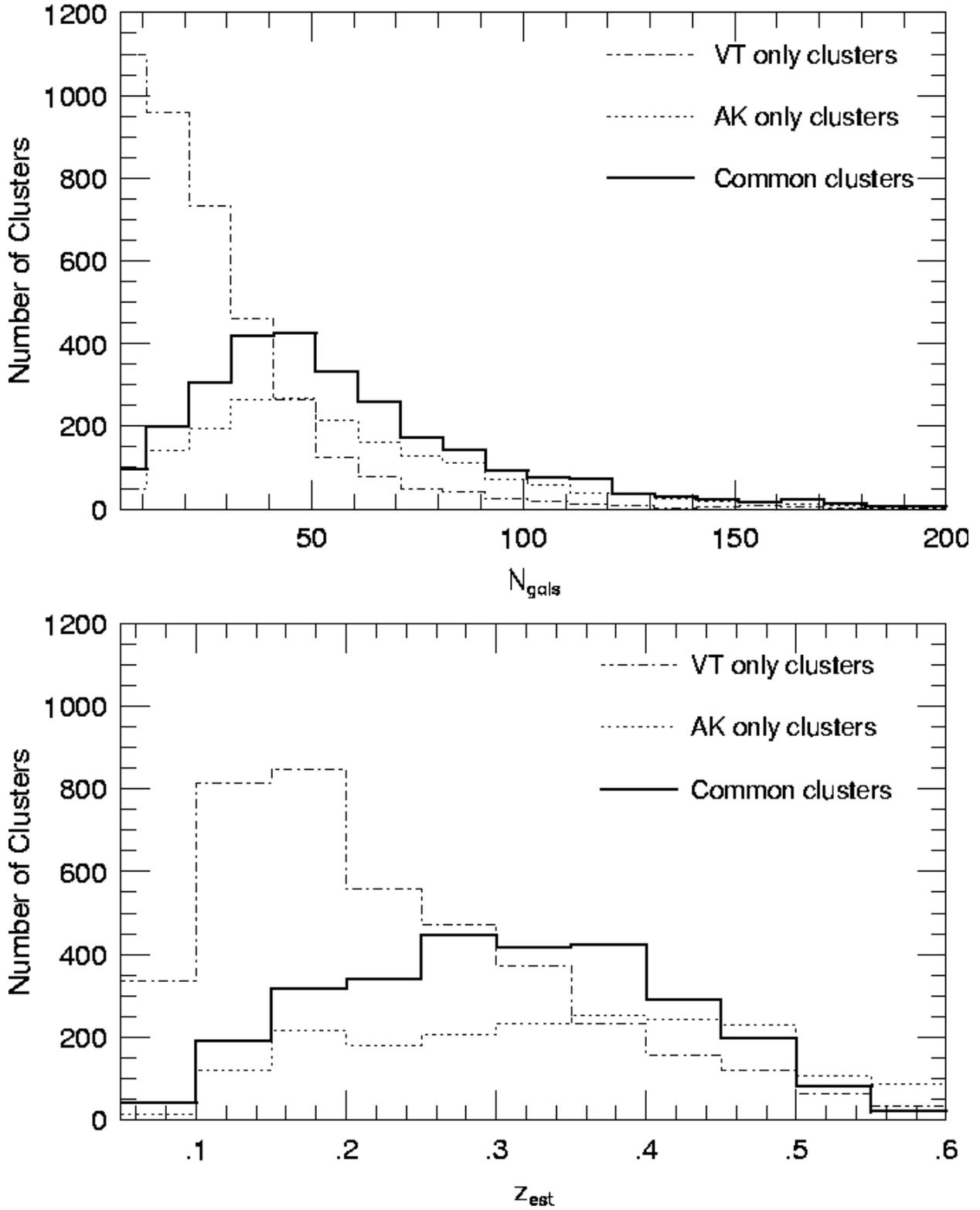}
\caption{{Richness (top) and estimated redshift (bottom) distributions for clusters
detected only by the VT code (dashed-dotted line), only by the AK code
(dotted line) and by both methods (heavy solid line).
\label{fig26}}}
\end{figure}

\clearpage

\begin{figure}
\plotone{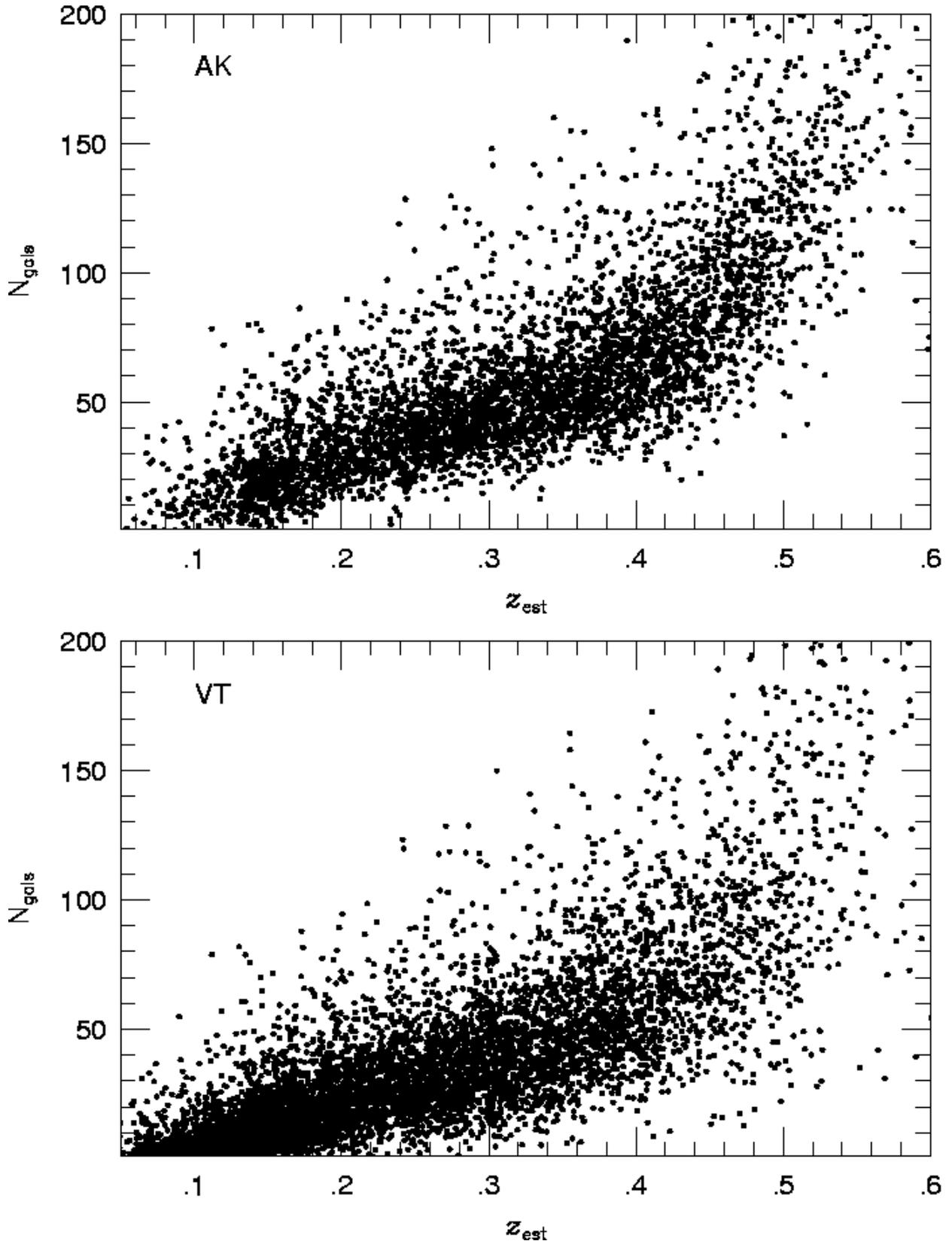}
\caption{{Richness {\it vs.} estimated redshift for the AK (top) and
VT (bottom) candidates.
\label{fig27}}}
\end{figure}

\clearpage

\begin{figure}
\plotone{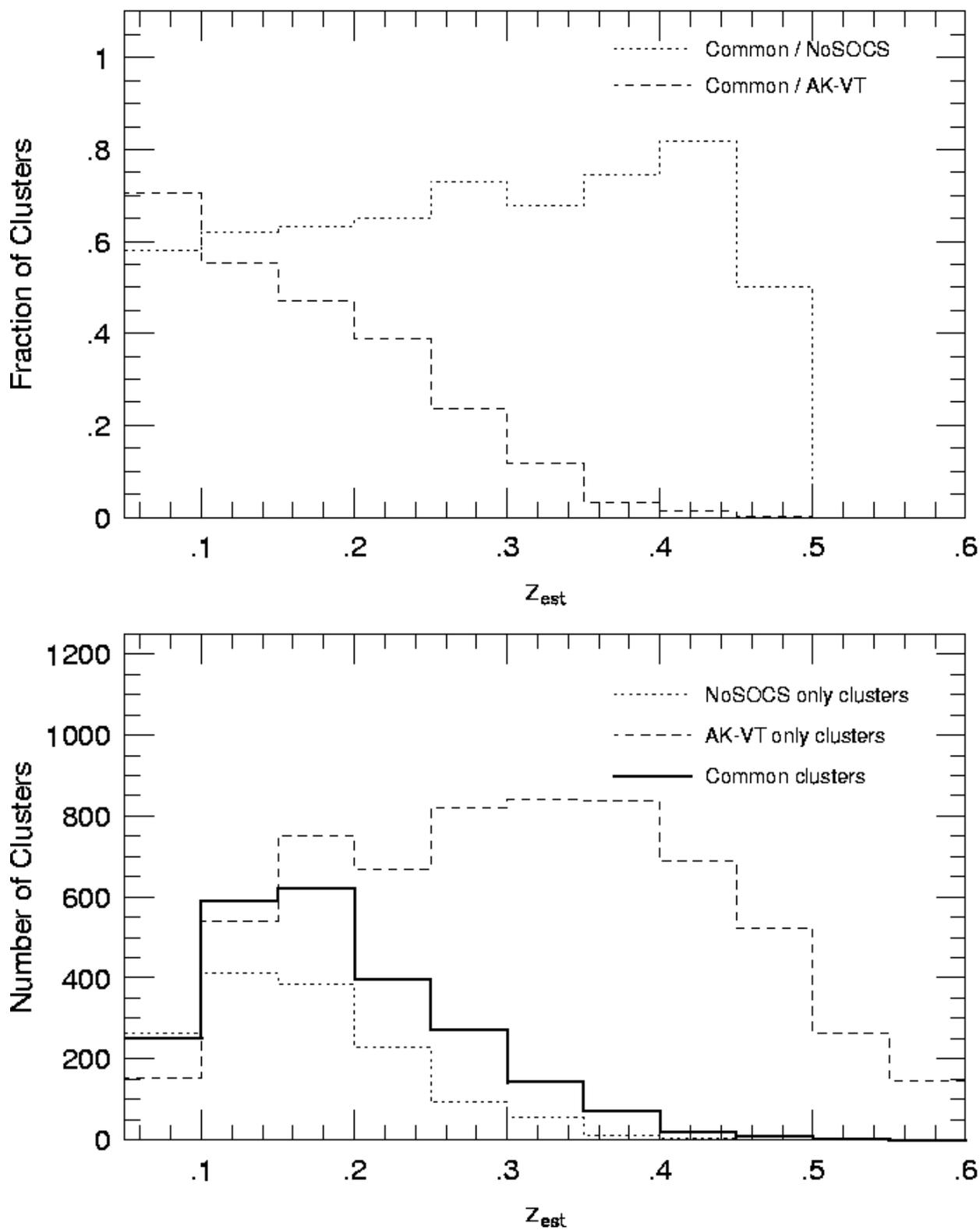}
\caption{{Estimated redshift distributions (bottom panel) for clusters
detected only in Papers II \& III (dotted line), only by this paper
(dashed line) and by both surveys (heavy solid line). The top
panel shows the ratio of common clusters to the NoSOCS catalog
(dotted line) and to the supplemental catalog presented here (dashed line).
\label{fig28}}}
\end{figure}

\clearpage

\begin{figure}
\plotone{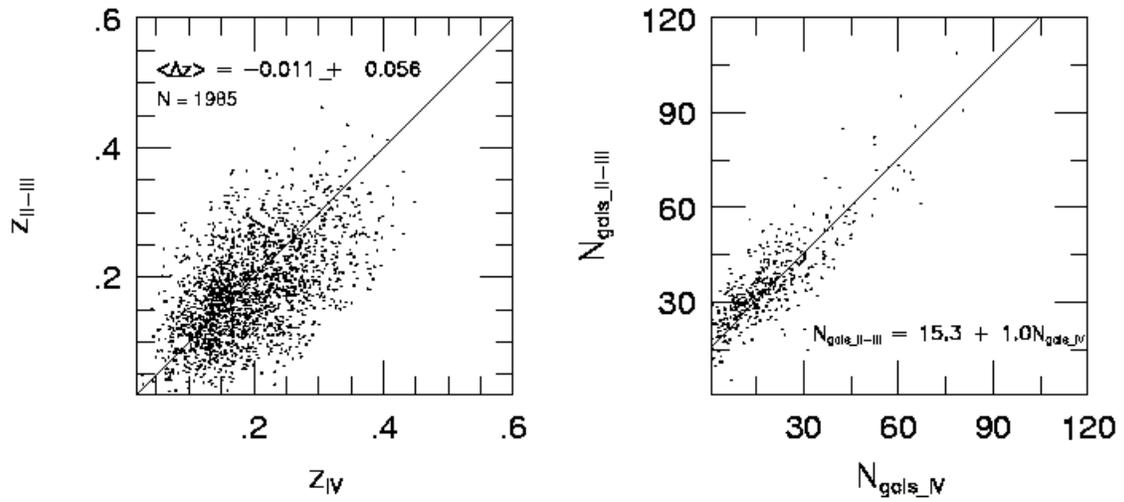}
\caption{{Comparison between our catalog and
those from Papers II and III. On both panels
the abscissa has the results for the current paper (IV),
while the ordinate shows the results from Papers II and III.
The left panel shows the redshift estimates for 1,985 common clusters
between this paper and the previous two. On the right richness
estimates for $z < 0.2$ clusters are compared.
 \label{fig29}}}
\end{figure}

\clearpage

\begin{figure}
\plotone{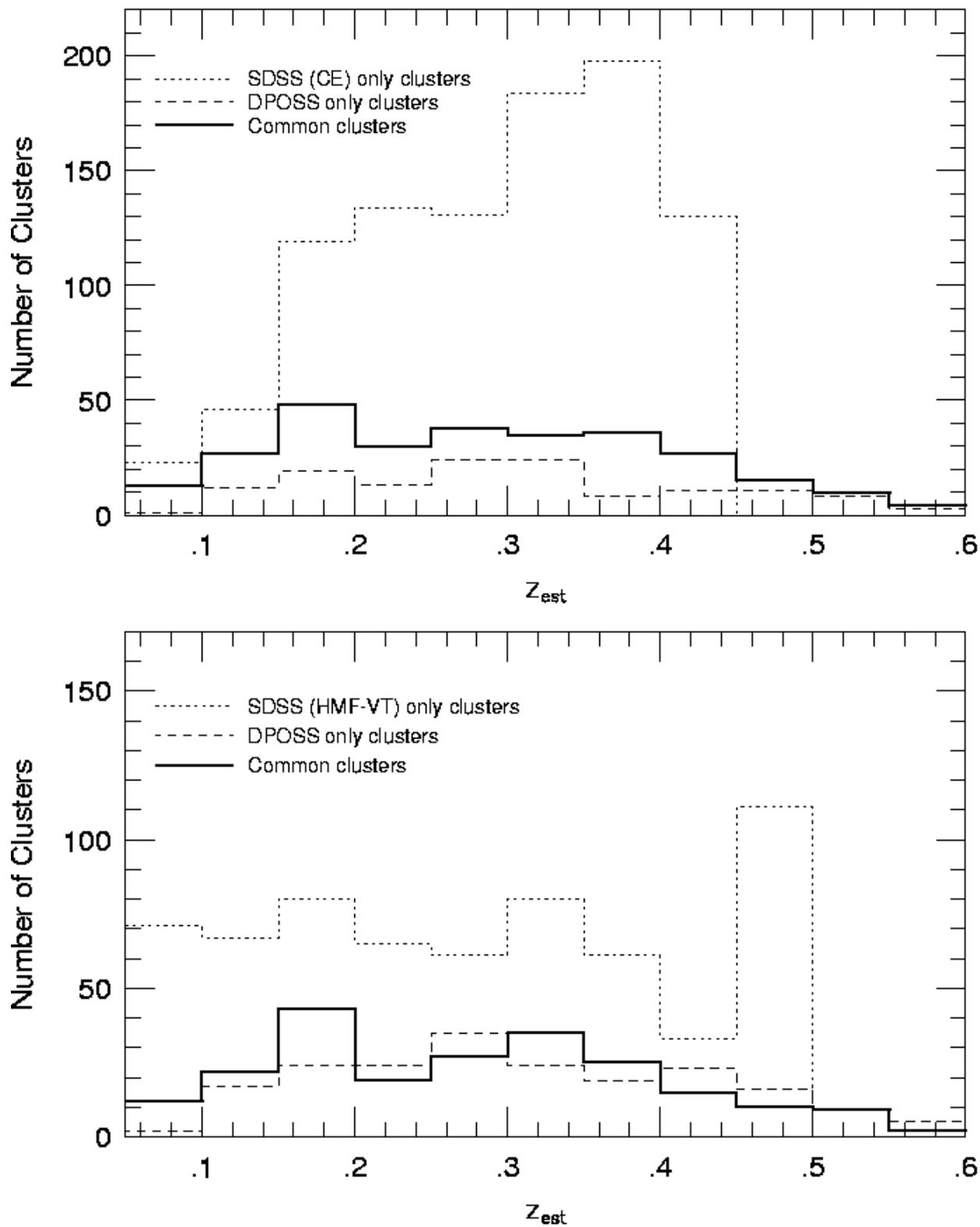}
\caption{Estimated redshift distributions of DPOSS only clusters
(dashed lines), common clusters (heavy solid line) and
SDSS only clusters (dotted line). In the bottom
panel DPOSS is compared to the catalog presented in
\citet{kim01}, while the comparison to the CE catalog
\citep{got02} is shown in the top panel. \label{fig30}}
\end{figure}

\clearpage

\begin{figure}
\plotone{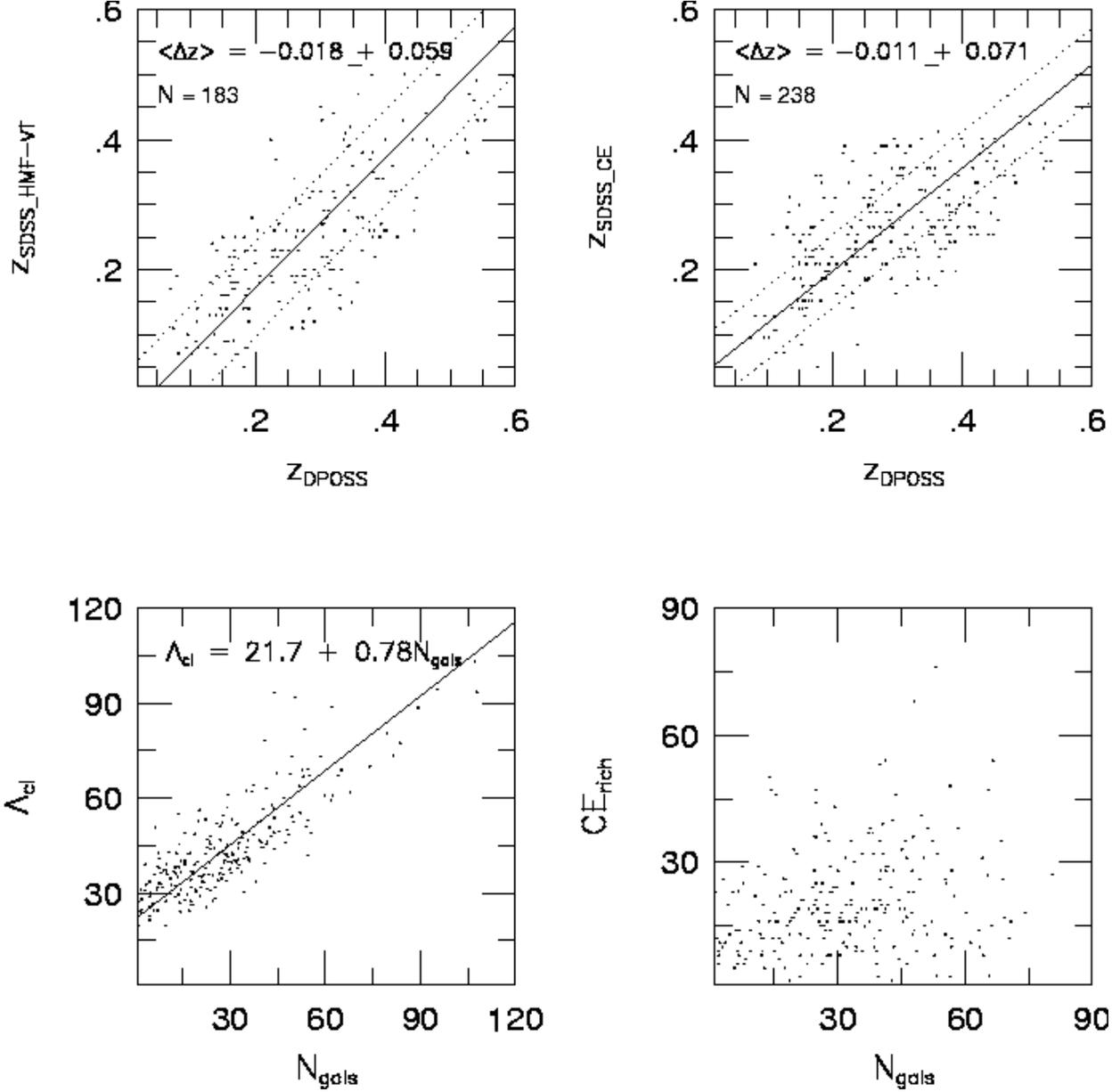}
\caption{Comparison between our catalog and
those from Kim (2001) \& Goto et al. (2002). In
the top panels we compare our redshift estimates
to those from \citet{kim01} (left) and \citet{got02} (right). 
The solid lines represent the best fit, while the
dashed lines give the combined $rms$ from the two catalogs, on each panel.
The bottom panels show the richness comparisons.
 \label{fig31}}
\end{figure}

\clearpage

\begin{figure}
\plotone{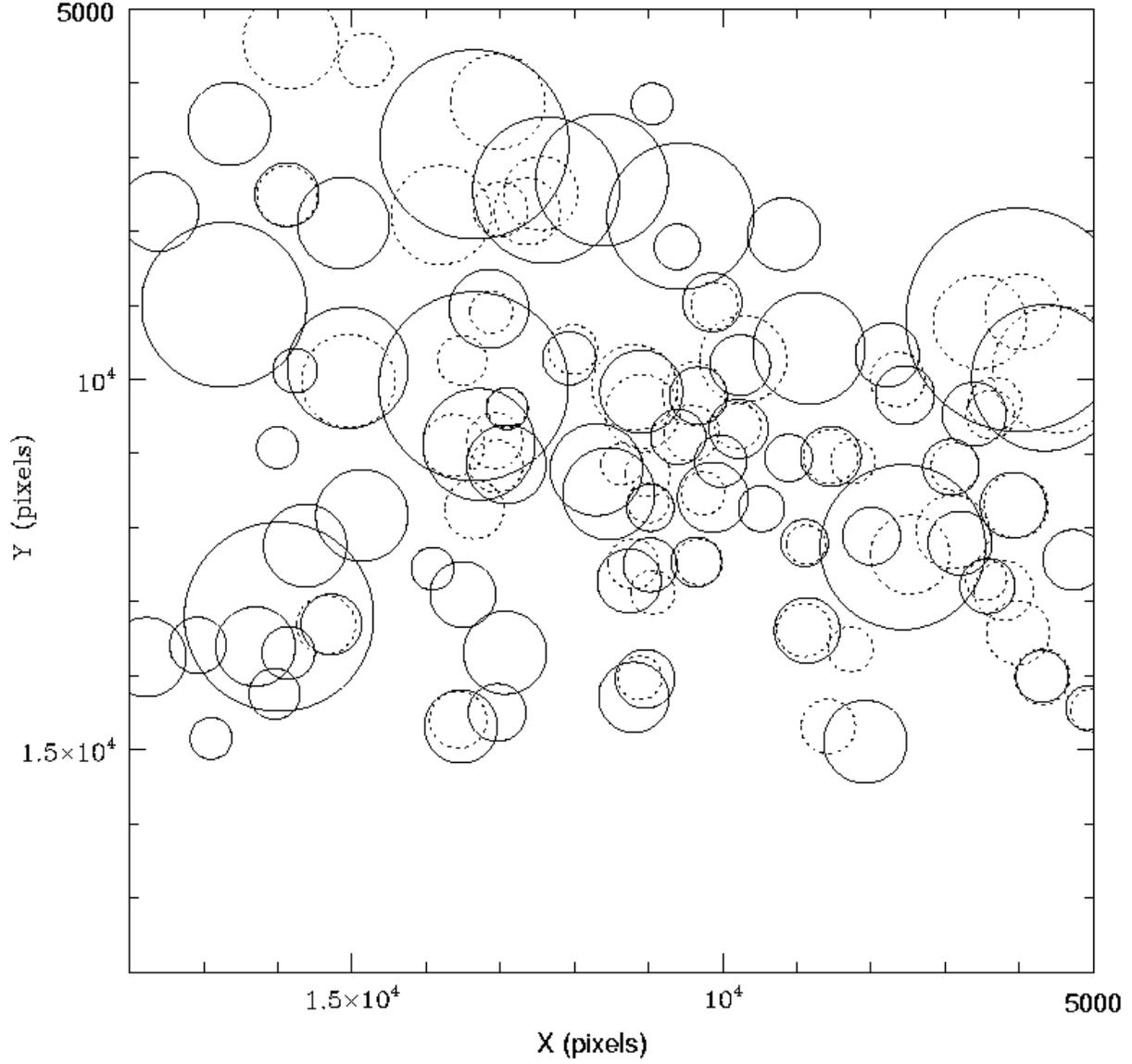}
\caption{The sky distribution of candidates detected in the central
region of DPOSS Field 824 ($\alpha = 5.65^{\circ}, ~\delta = 0.26^{\circ}$). 
AK-VT candidates are indicated by dashed circles
and the HMF-VT clusters by solid circles. We adopt a 1.5 $h^{-1}$ Mpc
radius for this illustration. When the redshift
estimator code fails we adopt $z = 0.3$ to draw the circle.
 \label{fig32}}
\end{figure}

\clearpage

\begin{deluxetable}{ccccccccl}
\tabletypesize{\footnotesize}
\tablecolumns{9}
\tablewidth{0pc}
\tablecaption{The Northern Sky Cluster Catalog Supplemental: Excerpt}
\tablehead{
\colhead{Name} & \colhead{RA (J2000.0)} & \colhead{Dec (J2000.0)} &
\colhead{$z_{phot-AK}$} & \colhead{$z_{phot-VT}$} & \colhead{$N_{gals-AK}$}
& \colhead{$N_{gals-VT}$} & \colhead{Plate} & \colhead{codes}
}
\startdata
NSCS J133609+292956 & 204.035858 & 29.498718 & 0.48 & 0.49 & 126.7 & 130.2 & 444 & AV \\
NSCS J132532+314613 & 201.384430 & 31.770241 & 0.44 & 0.45 &  67.2 &  67.7 & 444 & AV \\
NSCS J132421+310806 & 201.085632 & 31.134737 & 0.42 & 0.42 &  88.2 &  76.3 & 444 & AV \\
NSCS J132350+304404 & 200.959229 & 30.734278 & 0.46 & 0.49 & 128.8 & 158.0 & 444 & AV \\
NSCS J132049+320530 & 200.205231 & 32.091518 & 0.47 & 0.00 &  75.5 &   0.0 & 444 & A \\
NSCS J131747+315539 & 199.446884 & 31.927435 & 0.44 & 0.00 &  63.0 &   0.0 & 444 & A \\
NSCS J132036+314133 & 200.149994 & 31.692387 & 0.49 & 0.00 &  79.8 &   0.0 & 444 & A \\
NSCS J131644+312003 & 199.183975 & 31.334074 & 0.40 & 0.40 &  74.6 &  66.3 & 444 & AV \\
NSCS J133557+304355 & 203.987961 & 30.731905 & 0.42 & 0.42 &  81.3 &  76.7 & 444 & AV \\
NSCS J132303+303959 & 200.764053 & 30.666296 & 0.47 & 0.44 &  97.3 &  58.4 & 444 & AV \\
NSCS J132405+303250 & 201.019501 & 30.546940 & 0.43 & 0.44 &  70.4 &  71.5 & 444 & AV \\
NSCS J132336+302223 & 200.898285 & 30.373018 & 0.48 & 0.49 &  97.6 &  92.3 & 444 & AV \\
NSCS J134222+280715 & 205.593262 & 28.120775 & 0.44 & 0.44 & 217.4 & 201.3 & 445 & AV \\
NSCS J134316+282308 & 205.817719 & 28.385321 & 0.44 & 0.00 & 113.3 &   0.0 & 445 & A \\
NSCS J135107+305444 & 207.779678 & 30.912178 & 0.45 & 0.43 & 151.2 & 118.3 & 445 & AV \\
NSCS J140607+271029 & 211.527512 & 27.174681 & 0.49 & 0.50 & 120.5 & 131.5 & 446 & AV* \\
NSCS J140457+272800 & 211.235489 & 27.466606 & 0.50 & 0.50 & 159.7 & 133.1 & 446 & AV \\
NSCS J140612+282513 & 211.549500 & 28.420187 & 0.42 & 0.39 &  69.8 &  56.9 & 446 & AV \\
NSCS J140436+283443 & 211.148712 & 28.578556 & 0.45 & 0.46 & 175.6 & 150.9 & 446 & AV \\
NSCS J140614+282911 & 211.556107 & 28.486269 & 0.46 & 0.00 &  65.7 &   0.0 & 446 & A \\
NSCS J140733+283829 & 211.887100 & 28.641354 & 0.43 & 0.43 &  66.1 &  71.4 & 446 & AV \\
NSCS J141119+295227 & 212.827988 & 29.873974 & 0.46 & 0.00 & 134.0 &   0.0 & 446 & A \\
NSCS J143630+290031 & 219.126694 & 29.008404 & 0.49 & 0.42 & 132.4 &  83.3 & 447 & AV \\
NSCS J144045+292800 & 220.186127 & 29.466623 & 0.44 & 0.44 & 173.9 & 163.4 & 447 & AV \\
NSCS J143307+294227 & 218.279083 & 29.707354 & 0.41 & 0.42 &  91.5 &  97.2 & 447 & AV \\
NSCS J144332+315050 & 220.881851 & 31.847206 & 0.42 & 0.42 &  74.2 &  64.9 & 447 & AV \\
NSCS J142727+304153 & 216.864319 & 30.697830 & 0.47 & 0.00 & 126.6 &   0.0 & 447 & A \\
NSCS J150804+280926 & 227.017517 & 28.157183 & 0.48 & 0.00 & 167.3 &   0.0 & 448 & A \\
NSCS J145024+322408 & 222.598465 & 32.402100 & 0.44 & 0.00 &  99.6 &   0.0 & 448 & A * \\
NSCS J144940+321048 & 222.414612 & 32.179859 & 0.45 & 0.45 & 155.1 & 156.9 & 448 & AV \\
NSCS J150449+321740 & 226.201996 & 32.294426 & 0.45 & 0.00 & 109.6 &   0.0 & 448 & A \\
NSCS J150143+321237 & 225.430496 & 32.210087 & 0.45 & 0.00 & 176.5 &   0.0 & 448 & A \\
NSCS J145329+321024 & 223.369202 & 32.173214 & 0.45 & 0.50 & 134.0 & 198.4 & 448 & AV \\
NSCS J144802+320520 & 222.008209 & 32.088737 & 0.46 & 0.00 & 234.8 &   0.0 & 448 & A \\
NSCS J151043+320608 & 227.677841 & 32.102161 & 0.48 & 0.00 & 271.5 &   0.0 & 448 & A \\
\enddata
\end{deluxetable}


\begin{thebibliography}{}

\bibitem[Abell(1958)]{abe58} Abell, G.\ O.\ 1958, \apjs, 3, 211

\bibitem[Abell, Corwin \& Olowin(1989)]{aco89} Abell, G.\ O.,
Corwin, H.\ G.\ \& Olowin, R.\ P.\ 1989, \apjs, 70, 1

\bibitem[Annis et al.(1999)]{ann99} Annis, J. et al. 1999, \baas, 195

\bibitem[Arag\'on-Salamanca et al.(1993)]{ara93} Arag\'on-Salamanca, A.,
Ellis, R., Couch, W., Carter, D.\ 1993, \mnras, 262, 764

\bibitem[Bahcall \& Soneira(1983)]{bs83} Bahcall, N.\ A.\ \&
Soneira, R.\ M.\ 1983, \apj, 270, 20

\bibitem[Bahcall et al.(1997)]{bah97} Bahcall, N.\ A., Fan, X.,
Cen, R.\ 1997, \apj, 485, L53

\bibitem[Bahcall et al.(2003)]{bah03} Bahcall, N.\ A.\ \etal 2003, \apjs, 148, 243

\bibitem[Bertin \& Arnouts(1996)]{ber96} Bertin, E.~\&
Arnouts, S.\ 1996, \aaps, 117, 393

\bibitem[B{\"o}hringer et al.(2000)]{boh00} B{\"o}hringer,
H.\ \etal 2000, \apjs, 129, 435

\bibitem[Bramel, Nichol \& Pope(2000)]{bra00} Bramel, D.\ A.,
Nichol, R.\ C.\ \& Pope, A.\ C.\ 2000, \apj, 533, 601

\bibitem[Butcher \& Oemler(1984)]{bo84} Butcher, H. \& Oemler, A.\ 1984,
\apj, 285, 426

\bibitem[Carlberg et al.(1996)]{cal96} Carlberg, R., Yee, H., Ellingson, E., 
Abraham, R., Gravel, P., Morris, S., Pritchet, C. 1996, \apj, 462, 32

\bibitem[Carlberg et al.(1997)]{cal97} Carlberg, R., Morris, S.,
Yee, H., Ellingson, E.\ 1997, \apj, 479, L19

\bibitem[Coleman, Wu, \& Weedman(1980)]{col80} Coleman,
G.~D., Wu, C.~C., \& Weedman, D.~W.\ 1980, \apjs, 43, 393

\bibitem[Dalton, Maddox, Sutherland \& Efstathiou(1997)]{dal97}
Dalton, G.\ B., Maddox, S.\ J., Sutherland, W.\ J.\ \& Efstathiou, G.\ 1997,
\mnras, 289, 263

\bibitem[Djorgovski et al.(2004)]{djo04} Djorgovski, S.\ G., Gal, R.\ R., de
Carvalho, R.\ R., Odewahn, S.\ C., Mahabal, A., Brunner, R.\ J. \& Lopes,
P. 2004, \aj, {\it in prep.}

\bibitem[Doroshkevich et al.(1997)]{dor97} Doroshkevich, A., Gottlober, S.,
Madsen, S.\ 1997, \aaps, 123, 495

\bibitem[Dressler et al.(1997)]{dre97} Dressler, A., Oemler, A., Couch, W.,
Smail, I., et al.\ 1997, \apj, 490, 577

\bibitem[Ebeling \& Wiedenmann(1993)]{ebw93} Ebeling, H., Wiedenmann, G.\
1993, \pre, 47, 704

\bibitem[El-Ad et al.(1996)]{ela96} El-Ad, H., Piran, T. da Costa, L.\ 1996,
\apj, 462, 13

\bibitem[Evrard(1989)]{evr89} Evrard, A.\ 1989, \apj, 341, L71

\bibitem[Gal et al.(2000)]{gal00} Gal, R.\ R., de Carvalho,
R.\ R., Odewahn, S.\ C., Djorgovski, S.\ G.\ \& Margoniner, V.\ E.\ 2000,
\aj, 119, 12

\bibitem[Gal et al.(2003)]{gal03} Gal, R.\ R., de Carvalho, Lopes, P.\ A.\
A., Djorgovski, S.\ G., Brunner, R.\ J., Mahabal, A., Odewahn, S.\ C.\ 2003,
\aj, 125, 2064

\bibitem[Gal et al.(2004a)]{gal04a} Gal, R.\ R., de Carvalho, R.\ R.,
Odewahn, S.\ C., Djorgovski, S.\ G., Mahabal, A., Brunner, R.\ J., Lopes, P.
2004a, \aj, {\it accepted}

\bibitem[Gal et al.(2004b)]{gal04b} Gal, R.\ R., de Carvalho, Lopes, P.\ A.\
A., Djorgovski, S.\ G., Brunner, R.\ J., Mahabal, A., Odewahn, S.\ C.\ 2004b,
\aj, {\it in prep.}

\bibitem[Gilbank(2001)]{gil01} Gilbank, D.\ 2001, Ph.D.\ Thesis, University
of Durham

\bibitem[Gladders \& Yee(2000)]{gla00} Gladders, M.~D.~\&
Yee, H.~K.~C.\ 2000, \aj, 120, 2148

\bibitem[Gonzalez, Zaritsky, Dalcanton, \& Nelson(2001)]{gon01} Gonzalez,
A.~H., Zaritsky, D., Dalcanton, J.~J., \& Nelson, A.\ 2001, \apjs, 137, 117

\bibitem[Goto et al.(2002)]{got02} Goto, T.\ et al.\ 2002,
\aj, 123, 1807

\bibitem[Gunn, Hoessel \& Oke(1986)]{gun86} Gunn, J.\ E.,
Hoessel, J.\ G.\ \& Oke, J.\ B.\ 1986, \apj, 306, 30

\bibitem[Guzzo et al.(1992)]{guz92} Guzzo, L., Collins, C., Nichol, R.,
Lumsden, S.\ 1992, \apj, 393, 5

\bibitem[H$\phi$g et al.(2000)]{hog00} H$\phi$g, E., Fabricius, C.,
Makarov, V., Urban, S., Corbin, T., Wycoff, G., Bastian, U.,
Schwekendiek, P., Wicenec, A. 2000, \aap, 355, 27

\bibitem[Holden et al.(1999)]{hol99} Holden, B., Nichol, R., Romer, A.,
Metevier, A., Postman, M., Ulmer, M., Lubin, L.\ 1999, \aj, 118, 2002

\bibitem[Ikeuchi \& Turner(1991)]{ikt91} Ikeuchi, S.\ \& Turner, E.\ 1991,
\mnras, 250, 519

\bibitem[Kawasaki et al.(1998)]{kaw98} Kawasaki, W., Shimasaku, K., Doi, M.,
Okamura, S.\ 1998, \aaps, 130, 567

\bibitem[Kepner et al.(1999)]{kep99} Kepner, J., Fan, X.,
Bahcall, N., Gunn, J., Lupton, R.\ \& Xu, G.\ 1999, \apj, 517, 78

\bibitem[Kiang(1996)]{kia96} Kiang, T.\ 1996, Zeitschrift f{\" u}r
Astrophysik, 64, 433

\bibitem[Kim(2001)]{kim01} Kim, R.\ S.\ J.\ 2001, Ph.D.\ Thesis, Princeton

\bibitem[Kim et al.(2002)]{kim02} Kim, R.\ S.\ J.\ et al.\ 2002,
\aj, 123, 20

\bibitem[Lobo et al.(2000)]{lob00}
Lobo, C., Iovino, A., Lazzati, D.\ \& Chincarini, G.\ 2000, \aap, 360, 896

\bibitem[Lumsden, Nichol, Collins \& Guzzo(1992)]{lum92}
Lumsden, S.\ L., Nichol, R.\ C., Collins, C.\ A.\ \& Guzzo, L.\ 1992,
\mnras, 258, 1

\bibitem[Maddox, Efstathiou \& Sutherland(1996)]{mad96}
Maddox, S.\ J., Efstathiou, G.\ \& Sutherland, W.\ J.\ 1996, \mnras, 283,
1227

\bibitem[Margoniner et al.(2001)]{mar01} Margoniner, V. E., de 
Carvalho, R.\ R., Gal, R.\ R., Djorgovski, S.\ G.\ 2001, \apjl, 548, L143

\bibitem[Mullis et al.(2003)]{mul03} Mullis, C., McNamara, B., Quintana, H., 
Vikhlinin, A., Henry, J., Gioia, I., Hornstrup, A., Forman, W., Jones, C.
2003, \apj, 594, 154

\bibitem[Odewahn et al.(2004)]{ode04} Odewahn, S.\ C., Gal, R.\ R., de
Carvalho, R.\ R., Djorgovski, S.\ G., Mahabal, A., Brunner, R.\ J., Lopes,
P. A. A., Kohl Moreira, J. L., Stalder, B. 2004, \aj, {\it submitted}

\bibitem[Olsen et al.(1999)]{ols99} Olsen, L.~F.~et al.\ 1999, \aap, 345,
363

\bibitem[Paolillo et al.(2001)]{pao01} Paolillo, M., Andreon,
S., Longo, G., Puddu, E., Gal, R.~R., Scaramella, R., Djorgovski, S.~G., \&
de Carvalho, R.\ 2001, \aap, 367, 59

\bibitem[Picard(1991)]{pic91} Picard, A.\ 1991, Ph.D.\ Thesis, Caltech

\bibitem[Postman, Huchra \& Geller(1992)]{phg92} Postman, M., Huchra, J.,
\& Geller, M.\ 1992, \apj, 384, 404

\bibitem[Postman et al.(1996)]{pos96} Postman, M., Lubin, L.\ M., Gunn, J.\
E., Oke, J.\ B., Hoessel, J.\ G., Schneider, D.\ P.\ \& Christensen, J.\ A.\
1996, \aj, 111, 615

\bibitem[Postman et al.(1998)]{pos98} Postman, M., Lauer, T., Szapudi, I.,
Oegerle, W.\ 1998, \apj, 506, 33

\bibitem[Postman, Lauer, Oegerle, \& Donahue(2002)]{pos02} Postman, M.,
Lauer, T.~R., Oegerle, W., \& Donahue, M.\ 2002, \apj, 579, 93

\bibitem[Ramella et al.(2001)]{ram01} Ramella, M., Boschin, W., Fadda, D.,
Nonino, M.\ 2001, \aap, 368, 776

\bibitem[Reid et al.(1991)]{rei91} Reid, I.\ N.\ \etal \
1991, \pasp, 103, 661

\bibitem[Schuecker \& B{\"o}hringer(1998)]{sch98} Schuecker, P., B$\ddot
o$hringer, H.\ 1998, \aap, 339, 315


\bibitem[Shectman(1985)]{she85} Shectman, S.\ A.\ 1985,
\apjs, 57, 77

\bibitem[Shewchuk(1996)]{shw96} Shewchuk, J.\ 1996, in First Workshop on
Applied Comptutational Geometry, ACM

\bibitem[Silverman(1986)]{sil86} Silverman, B.\ W.\ 1986,
{\it Monographs on Statistics and Applied Probability}, London: Chapman and
Hall

\bibitem[Squires et al.(1996)]{squ96} Squires, G., Kaiser,
N., Babul, A., Fahlman, G., Woods, D., Neumann, D.~M., \& B$\ddot o$hringer,
H.\ 1996, \apj, 461, 572

\bibitem[Struble \& Rood(1999)]{str99} Struble, M.\ F.\ \&
Rood, H.\ J.\ 1999, \apjs, 125, 35

\bibitem[Sunyaev \& Zeldovich(1980)]{sun80} Sunyaev, R. A. \& Zeldovich, 
Y. B.\ 1980, \araa, 18, 537

\bibitem[Tyson \& Fischer(1995)]{tys95} Tyson, J.~A.~\&
Fischer, P.\ 1995, \apjl, 446, L55

\bibitem[Viana \& Liddle(1996)]{via96} Viana, P.\ T.\ P.,
Liddle, A.\ R.\ 1996, \mnras, 281, 323

\bibitem[Vikhlinin et al.(1998)]{vil98} Vikhlinin, A., McNamara, B, Forman,
W., Jones, C., Quintana, H., Hornstrup, A.\ 1998, \apj, 502, 558

\bibitem[Wittman et al.(2001)]{wit01} Wittman, D., Tyson, J. A., 
Margoniner, V. E., Cohen, J. G., Dell'Antonio, I. P.\ 2001, \apj, 557, 89

\bibitem[Zaninetti(1995)]{zan95} Zaninetti, L.\ 1995, \aaps, 109, 71

\bibitem[Zwicky, Herzog \& Wild(1968)]{zwi68} Zwicky, F.,
Herzog, E.\ \& Wild, P.\ 1968, Catalogue of Galaxies and of Clusters of
Galaxies, Pasadena: California Institute of Technology (CIT), 1961-1968
\end{thebibliography}
\end{document}